\documentclass[twocolumn,english,prb,showpacs]{revtex4-1}
\usepackage{epsfig}
\usepackage{graphicx}
\usepackage{amsfonts}
\usepackage[figuresright]{rotating}
\usepackage{amssymb}
\usepackage{amsmath}
\usepackage{amsmath}
\usepackage{graphicx}
\usepackage{hyperref}
\usepackage{color}
\usepackage{soul}
\usepackage{tabularx}
\usepackage{array}

\newcolumntype{P}[1]{>{\centering\arraybackslash}p{#1}}

\usepackage{amsmath,amssymb,amsfonts}	
\usepackage{color}
\usepackage{amsmath}
\usepackage{amsfonts}
\usepackage{amssymb}
\usepackage{graphicx}
\usepackage{slashed}            
\usepackage{bbm}                
\usepackage{xspace}				

\usepackage{tikz}
\usetikzlibrary{arrows,shapes}
\usetikzlibrary{trees}
\usetikzlibrary{matrix,arrows} 				
\usetikzlibrary{positioning}				
\usetikzlibrary{calc,through}				
\usetikzlibrary{decorations.pathreplacing}  
\usepackage{pgffor}							

\usetikzlibrary{decorations.pathmorphing}	
\usetikzlibrary{decorations.markings}
\tikzset{
    vector/.style={decorate, decoration={snake}, draw},
	provector/.style={decorate, decoration={snake,amplitude=2.5pt}, draw},
	antivector/.style={decorate, decoration={snake,amplitude=-2.5pt}, draw},
    fermion/.style={draw=black, postaction={decorate},
        decoration={markings,mark=at position .55 with {\arrow[draw=black]{>}}}},
    fermionbar/.style={draw=black, postaction={decorate},
        decoration={markings,mark=at position .55 with {\arrow[draw=black]{<}}}},
    fermionnoarrow/.style={draw=black},
    gluon/.style={decorate, draw=black,
        decoration={coil,amplitude=4pt, segment length=5pt}},
    scalar/.style={dashed,draw=black, postaction={decorate},
        decoration={markings,mark=at position .55 with {\arrow[draw=black]{>}}}},
    scalarbar/.style={dashed,draw=black, postaction={decorate},
        decoration={markings,mark=at position .55 with {\arrow[draw=black]{<}}}},
    scalarnoarrow/.style={dashed,draw=black},
    electron/.style={draw=black, postaction={decorate},
        decoration={markings,mark=at position .55 with {\arrow[draw=black]{>}}}},
	bigvector/.style={decorate, decoration={snake,amplitude=4pt}, draw},
}

\tikzstyle{block} = [draw, rectangle, 
    minimum height=3em, minimum width=6em]

\usepackage{hyperref}			

\begin{document}

\title{Thermoelectric transport in double-Weyl semimetals}
\author{Qi Chen}
\email[]{chenqi0805@gmail.com}
\author{Gregory A. Fiete}
\affiliation{Department of Physics, The University of Texas at
Austin, Austin, TX, 78712, USA}
\begin{abstract}

We study the thermoelectric properties of a double-Weyl fermion system, possibly realized in $\mathrm{HgCr_2Se_4}$ and $\mathrm{SrSi_2}$,  by a semi-classical Boltzmann transport theory. We  investigate different relaxation processes including short-range disorder and electron-electron interaction on the thermoelectric transport coefficients. It is found that the anisotropy of the band dispersion for in-plane and out-of-plane momentum directions affects the relaxation time for transport in different directions. The transport also exhibits an interesting directional dependence on the chemical potential and model parameters, differing from a simple isotropic quadratic or linearly dispersing electron gas. By applying a static magnetic field along the linearly dispersing direction, the longitudinal and transverse electrical and thermal magneto-conductivity show a similar dependence on the in-plane cyclotron frequency to the linear dispersing Weyl nodes. By including internode scattering, we find that the chiral anomaly contribution to the thermoelectric coefficients doubles that of a linearly dispersing Weyl node in both the semi-classical and quantum regimes. A magnetic field applied along the quadratically dispersing direction will split the double-Weyl point into two single-Weyl points with the same chirality.
\end{abstract}
\maketitle

\section{Introduction}

In recent years, topological semimetals have enjoyed a surge of interest.\cite{Vafek:arcm14}   In these systems the isolated Fermi surfaces or Fermi points support gapless quasiparticle excitations only in the vicinity of isolated band touching points in the Brillouin zone (BZ). Well-known experimental examples of the semimetals include linearly dispersing massless Dirac quasiparticles in both two dimensions (2D) and three dimensions (3D). For example, monolayer graphene in 2D\cite{Novoselov22102004, Novoselov2005, Zhang2005} and $\mathrm{Bi_{1-x} Sb_x}$\cite{Lenoir1996, Ghosalprb2007, Castrormp2009}, $\mathrm{Pb_{1-x}Sn_xTe}$\cite{Dornhaus1983, Goswamiprl2011}, and $\mathrm{Cd_3As_2}$\cite{Borisenkoprl2014}, $\mathrm{Na_3Bi}$\cite{Liu21022014} in 3D. It is also possible to realize parabolic semimetals which possess parabolic dispersions at band touching, e.g., such as Bernal-stacked bilayer graphene,\cite{Novoselov2006} [111] grown LaNiO$_3$ bilayers\cite{Yang:prb11a,Ruegg11_2,Ruegg_Top:prb13,Ruegg:prb12}, and [111] grown $\mathrm{Y_2Ir_2O_7}$ films \cite{Hu:sr15} in 2D, and $\mathrm{HgTe},$\cite{Dornhaus1983} gray tin,\cite{Grovesprl1963} and the normal state at high temperatures for some $227$ irradiates, such as $\mathrm{Pr_2Ir_2O_7}$\cite{Nakatsujiprl2006, Machida2009, Abrikosov1974, Moonprl2013, Herbutprl2014, Hsin-Hua2014} in 3Ds. Dirac (Weyl) semimetals have linearly dispersing excitations [which obey the 3D Dirac (Weyl) equation] from degenerate band touching points referred to as Dirac (Weyl) nodes. For Weyl semi-metals, electronic states around the band degeneracy points possess a nonzero Berry curvature, which gives rise to nontrivial momentum space topology.\cite{Vafek:arcm14}

In addition to Weyl semimetals, a new 3D topological semimetal, termed the double-Weyl semimetal, has been proposed in materials with certain point-group symmetries.\cite{Fangprl2012, ShinMing2015} The double-Weyl semimetals have band touching points with quadratic dispersions in two directions, e.g., the $\hat{x}$-$\hat{y}$ plane, and linear dispersion in the third direction, e.g. the $\hat{z}$ direction. The double-Weyl nodes are protected by $C_4$ or $C_6$ rotation symmetry and are predicted to be realized in the 3D semimetal $\mathrm{HgCr_2Se_4}$ in the ferromagnetic phase, with a pair of double-Weyl nodes along the $\Gamma Z$ direction\cite{Fangprl2012, Xuprl2011}, as well as in $\mathrm{SrSi_2}$.\cite{ShinMing2015}.  Transport experiments\cite{Guanprl2015} in $\mathrm{HgCr_2Se_4}$ confirm the half-metallic property of $\mathrm{HgCr_2Se_4}$, in qualitative agreement with theory. The (anti-) double-Weyl node possesses a monopole charge of $(-2) + 2$, and the double-Weyl semimetal shows double-Fermi arcs on the surface BZ\cite{Fangprl2012, Xuprl2011, ShinMing2015}. In $\mathrm{SrSi_2}$, Weyl nodes with opposite charges are located at different energies due to the absence of mirror symmetry\cite{ShinMing2015}.  A one-loop renormalization group study\cite{Hsin-Huaprb2015, Shao-Kaiprb2015} of interacting double-Weyl fermions shows that there exists a stable fixed point at which the long-range Coulomb interaction is screened anisotropically.

Although there has been much work on the electrical and thermal transport properties of Dirac and semi-Dirac fermions in 2D\cite{Gusyninprl2005, Gusynin2006, GusyninJPCM2007, Fritzprb2008, Castrormp2009, Novoselov2006, IgorJPSJ2013, Schuttprb2011, Peres20061559, Adroguer2015} and Weyl fermions in 3D\cite{Nielsen1983, Burkovprb2011, Fradkinprb1986I, Fradkinprb1986II, Hosurprl2012, Ajiprb2012, LoganayagamJHEP2012, Ominatoprb2014, Rosensteinprb2013, Jhoprb2013, Huangprb2013, Gorbarprb2014, HuangNJP2013, Burkovprb2014, Sbierskiprl2014, Landsteinerprb2014, Syzranovprl2015, Nandkishoreprb2014, Kobayashiprl2014, Biswasprb2014, Rexprb2014,Song2015, Luprb2015} (with a particular focus on the case when the Fermi energy lies at the Dirac or Weyl node), a thermoelectric transport study of the double-Weyl fermion system in 3D is still lacking. In this work, we investigate the electronic contribution to the electrical and thermal conductivity and thermopower of double-Weyl semimetals. The semiclassical Boltzmann equation is applied to analytically calculate the thermoelectric coefficients for noninteracting electrons, as well as for relaxation from electron-electron interactions. Moreover, the diffusive transport coefficients are calculated by considering short-range disorder within first Born approximation. For diffusive transport, we find good agreement in electrical conductivity between Boltzmann and diagrammatic approaches. We also investigated the effect of a static magnetic field along the $\hat{z}$ direction (linearly dispersing direction), which preserves the double Weyl node (a magnetic field applied in the $xy$-plane splits the double Weyl node into two single Weyl nodes with the same chirality). For a magnetic field perpendicular to the electric field or temperature gradient, the transverse electrical and thermal conductivity is linear in magnetic field strength in the weak field regime, while the longitudinal components have a quadratic field dependence that decreases the magnitude of the thermal conductivity, similar to the case of a single Weyl point\cite{Rexprb2014}. For a magnetic field in parallel with the electric field or temperature gradient, the longitudinal electrical or thermal conductivity shows a positive quadratic dependence on the magnetic field strength, which is the characteristic feature of Weyl semimetals.  We also investigated the chiral anomaly effect related to the internode scattering rate. Similar to the linearly dispersing Weyl node, the thermoelectric coefficients depend on the magnetic field strength quadratically in the semi-classical regime and linearly in the quantum regime\cite{Rexprb2014, DTSonprb2013}.

Our paper is organized as follows. In Sec. \ref{sec: no field}, the transport coefficients are obtained by a semiclassical approach for noninteracting electrons, relaxation from electron-electron interactions, and short-range disorder. In Sec. \ref{sect: with field}, the effect of a static magnetic field is investigated on thermoelectric transport properties. In Sec. \ref{sect:summary}, we present conclusions and a discussion of our results. A comparison to diagrammatic result in diffusive transport is addressed in the Appendix.

\section{THERMOELECTRIC TRANSPORT QUANTITIES}
\label{sec: no field}

The non-interacting low energy effective Hamiltonian for a single double-Weyl fermion system is given by,
\begin{equation}
\label{eq:freeham}
H^{\chi}=\frac{k_x^2-k_y^2}{m}\sigma _x+\frac{2k_xk_y}{m}\sigma _y+\chi v_zk_z\sigma _z,
\end{equation}
where $m$ is the effective mass in $\hat{x}$-$\hat{y}$ plane, $v_z$ is the Fermi velocity in the $\hat{z}$ direction, and $\chi=\pm 1$ represents the chirality. The spectrum of Eq.(\ref{eq:freeham}),
\begin{equation}
\epsilon _{\pm}(\mathbf{k})=\pm \sqrt{\left(\frac{k_x^2+k_y^2}{m}\right)^2+\left(v_z k_z\right)^2},
\end{equation}
is linear in $\hat{z}$ direction and quadratic in $\hat{x}$-$\hat{y}$ plane. The density of states,
\begin{equation}
g(\epsilon )=\frac{m}{\hbar
^3v_z}\left|\frac{\epsilon }{4\pi }\right|,
\end{equation}
has a linear dependence on energy. In our work, we only consider the contribution to the thermoelectric coefficients from the electronic degrees of freedom, assuming the temperature is low-enough that phonons make a negligible contribution. Due to the appearance of a ferromagnetic phase in $\mathrm{HgCr_2Se_4}$\cite{Fangprl2012, Xuprl2011}, the magnon contribution to the thermal transport might become important, but will be left for future study.  Our theory should be directly applicable to systems without broken time-reversal symmetry, as well as those with broken time-reversal symmetry where the itinerant motion of the electrons is the dominant contribution to the thermal transport.

\subsection{Boltzmann transport theory}

In order to calculate thermoelectric properties, we introduce the semiclassical Boltzmann formalism, following the notation of Ref. [\onlinecite{Rexprb2014}]. The electrical current for a given chirality $\chi$ of a double Weyl node is, 
\begin{widetext}
\begin{eqnarray}
\mathbf{J}^{\chi }&=&-e\int d\mathbf{k}f^{\chi }(\mathbf{k},\mathbf{r},t)\dot{\mathbf{r}} \nonumber \\ 
&&-\nabla \times \left[\frac{1}{\beta (\mathbf{r})}\int d\mathbf{k}\mathbf{ }\frac{e}{\hbar
}\mathbf{\Omega }^{\chi }(\mathbf{k})\text{Log}[1+\exp (-\beta (\mathbf{r})(\epsilon_{\bf k} -\mu ))]\right] \nonumber \\
&=&-e\int d\mathbf{k}\mathbf{ }f^{\chi }(\mathbf{k},\mathbf{r},t)\left(\frac{1}{\hbar }\frac{\partial \epsilon_{\bf k}}{\partial \mathbf{k}}+\frac{e}{\hbar }\mathbf{E}\times
\mathbf{\Omega }^{\chi }(\mathbf{k})\right) \nonumber \\
&&+\left(-\frac{\nabla T}{T}\right)\times \frac{e}{\hbar } \left( \int d\mathbf{k} \mathbf{\Omega }^{\chi }(\mathbf{k})k_BT
\text{Log}[1+\exp (-\beta (\mathbf{r})(\epsilon_{\bf k} -\mu ))]+\int d\mathbf{k} \mathbf{\Omega }^{\chi }(\mathbf{k})(\epsilon_{\bf k} -\mu )f_{eq} \right),
\end{eqnarray}
\end{widetext}
where $e$ is the electrical charge, $\hbar$ is Planck's constant devided by $2\pi$, $\mu$ is the chemical potential, $\epsilon_{\bf k}$ is the dispersion for quasiparticles with wave vector $\mathbf{k}$, $f^{\chi}$ is the non-equilibrium distribution function for electrons at the double Weyl point with chirality $\chi$ and $f_{eq}$ is the equilibrium Fermi-Dirac distribution. Here $\mathbf{\Omega}^{\chi}=\nabla_{\mathbf{k}}\times \mathbf{A}^{\chi}(\mathbf{k})$ is the Berry curvature and $\mathbf{A}^{\chi}(\mathbf{k})=i \langle u^{\chi}_k|\nabla_{\mathbf{k}}|u^{\chi}_k \rangle$ is the Berry connection associated with the Bloch state $|u^{\chi}_k \rangle$.  The heat current is given by
\begin{widetext}
\begin{eqnarray}
\mathbf{J}^{Q, \chi }&=&\mathbf{J}^{E, \chi }-\mu \mathbf{J}^{\chi } \nonumber \\
&=&\int d\mathbf{k}\mathbf{\text{  }}f^{\chi }(\mathbf{k},\mathbf{r},t)(\epsilon_{\bf k} -\mu ) \left(\frac{1}{\hbar }\frac{\partial \epsilon_{\bf k}}{\partial \mathbf{k}}+\frac{e}{\hbar
}\mathbf{E}\times \mathbf{\Omega }^{\chi }(\mathbf{k})\right) \nonumber \\
&&+\mathbf{E}\times k_BT\int d\mathbf{k}\mathbf{ }\frac{e}{\hbar }\mathbf{\Omega }^{\chi }(\mathbf{k})\text{Log}[1+\exp
(-\beta (\epsilon_{\bf k} -\mu ))]-\nabla T\times \frac{\partial \mathbf{M}_Q}{\partial T},
\end{eqnarray}
\end{widetext}
with the heat magnetization defined as $\mathbf{M}_Q \equiv \mathbf{M}_E-\mu  \mathbf{M}_N$\cite{Cooperprb1997, TaoQinprl2011}. The explicit expressions for $\mathbf{M}_E$ and $\mathbf{M}_N$ are cumbersome and given in Ref. [\onlinecite{Cooperprb1997, DiXiaoprl2006, TaoQinprl2011}].   The non-equilibrium distribution function \(f^{\chi }(\mathbf{k}, \mathbf{r}, t)\) is determined by the Boltzmann transport equation,
\begin{widetext}
\begin{eqnarray}
\label{eq:Boltzmann}
\frac{\partial f^{\chi }(\mathbf{k},\mathbf{r},t)}{\partial t}+\left\{\mathbf{\dot{r}} \cdot \frac{\partial f^{\chi
}(\mathbf{k},\mathbf{r},t)}{\partial \mathbf{r}}+\mathbf{\dot{k}}\cdot \frac{\partial f^{\chi }(\mathbf{k},\mathbf{r},t)}{\partial \mathbf{k}}\right\}=I_{\text{coll}}^{\chi }\left\{f^{\chi }(\mathbf{k},\mathbf{r},t)\right\},
\end{eqnarray}
with
\begin{eqnarray}
\label{eq: velocityandacceleration}
\dot{\mathbf{r}}&=&\left(1+\frac{e}{c}\mathbf{B}\mathbf{\cdot }\mathbf{\Omega }^{\chi }(\mathbf{k})\right)^{-1}\left[\mathbf{v}_{\mathbf{k}}+e\mathbf{E}\times \mathbf{\Omega
}^{\chi }(\mathbf{k})+\frac{e}{c}\left(\mathbf{\Omega }^{\chi }(\mathbf{k})\cdot \mathbf{v}_{\mathbf{k}}\right)\mathbf{B}\right], \nonumber \\
\dot{\mathbf{p}}&=&\left(1+\frac{e}{c}\mathbf{B}\mathbf{\cdot }\mathbf{\Omega }^{\chi }(\mathbf{k})\right)^{-1}\left[e\mathbf{E}+\frac{e}{c}\mathbf{v}_{\mathbf{k}}\times \mathbf{B}+\frac{e^2}{c}(\mathbf{E\cdot
B})\mathbf{\Omega }^{\chi }(\mathbf{k})\right],
\end{eqnarray}
\end{widetext}
where $\mathbf{v}_{\mathbf{k}}=\frac{1}{\hbar} \frac{\partial \epsilon_{\bf k}}{\partial \mathbf{k}}$ is the group velocity. The right hand side of Eq.(\ref{eq:Boltzmann}) is a collision integral. If we only consider intra-node scattering (chiral anomaly effects from inter-node scattering will be addressed in a later section) in relaxation time approximation\cite{Mermin1976}, then
\begin{equation}
I_{\text{coll}}^{\chi }\left\{f^{\chi }(\mathbf{k},\mathbf{r},t)\right\}=-\frac{f^{\chi }(\mathbf{k},\mathbf{r},t)-f_{\text{eq}}}{\tau_{\rm intra} }.
\end{equation}
The thermoelectric response is obtained from\cite{Mermin1976}
\begin{eqnarray}
J_{\alpha }&=&L_{\alpha \beta }^{11}E_{\beta }+L_{\alpha \beta }^{12}\left(-\nabla _{\beta }T\right),\nonumber \\
J_{\alpha }^Q&=&L_{\alpha \beta }^{21}E_{\beta }+L_{\alpha \beta }^{22}\left(-\nabla _{\beta }T\right),
\end{eqnarray}
with summation over repeated Greek indices. The longitudinal thermoelectric coefficients are given by\cite{Mermin1976}
\begin{eqnarray}
L_{\alpha \alpha }^{11}&=&\sigma _{\alpha \alpha }=\mathcal{L}_{\alpha }^0, \nonumber \\
L_{\alpha \alpha }^{21}&=&T L_{\alpha \alpha }^{12}=\frac{-\mathcal{L}_{\alpha }^1}{e}, \nonumber \\
L_{\alpha \alpha }^{22}&=&\frac{\mathcal{L}_{\alpha \alpha }^2}{e^2T},
\end{eqnarray}
with
\begin{eqnarray}
\mathcal{L}_{\alpha }^n&=&e^2\sum_{s=\pm}\int \frac{d^3\mathbf{k}}{(2\pi )^3} \tau \left(\epsilon_{s\bf k})\right)\left(-\frac{\partial f}{\partial \epsilon _{s\bf k}}\right)\left(\frac{1}{\hbar }\frac{\partial \epsilon_{s\bf k}}{\partial k_{\alpha}}\right)^2 \times \nonumber \\
&&\left(\epsilon_{s\bf k}-\mu \right)^n,
\end{eqnarray}
where $s=\pm$ is the band index. The thermal conductivity and the Seebeck coefficient (thermopower) are defined as,
\begin{equation}
\label{eq:kappa}
\kappa _{\alpha  \beta }=L_{\alpha  \beta }^{22}-L_{\alpha  \gamma }^{21}\left(L_{\gamma  \rho }^{11}\right)^{-1}L_{\rho  \beta }^{12},
\end{equation}
\begin{equation}
S_{\alpha }=\frac{L_{\alpha  \alpha }^{12}}{L_{\alpha  \alpha }^{11}}.
\end{equation}
The anomalous transverse response is given by,\cite{DiXiaoprl2006, TaoQinprl2011} 
\begin{eqnarray}
\label{eq:L_anomalous}
L_{\alpha \beta }^{11}&=&\sigma _{\alpha \beta }(\mu)=-\epsilon _{\alpha \beta \gamma }\frac{e^2}{\hbar } \sum_s \int \frac{d^3\mathbf{k}}{(2\pi )^3}\Omega _{\gamma
}^{\chi }(\mathbf{k})f_{s,\mathbf{k}}, \nonumber \\
L_{\alpha \beta }^{12}&=&T L_{\alpha \beta }^{21}=-\frac{1}{e}\int d\epsilon \frac{\partial f(\epsilon)}{\partial \mu }\sigma _{\alpha \beta }(\epsilon )\frac{\epsilon -\mu }{T} ,\nonumber \\
L_{\alpha \beta }^{22}&=&\frac{1}{e^2T}\int d\epsilon \left(-\frac{\partial f(\epsilon)}{\partial \epsilon }\right)(\epsilon -\mu )^2\sigma _{\alpha \beta }(\epsilon).
\end{eqnarray}
\
\
\
\subsection{Free electron}

\subsubsection{Longitudinal thermoelectric properties}

As a first step, we use the semiclassical approach to calculate the dc conductivity evaluated by the Kubo formula in Ref. [\onlinecite{Hsin-Huaprb2015}] by assuming an energy and momentum independent scattering time:
\begin{widetext}
\begin{eqnarray}
\sigma _{x x}^{\text{dc}}=\sigma _{y y}^{\text{dc}}&=&\frac{m e^2}{3\pi  m v_z\hbar ^3}\frac{\beta }{4}\int d\epsilon  \epsilon ^2\left(\text{sech}^2\left(\frac{\beta (\epsilon +\mu )}{2}\right)+\text{sech}^2\left(\frac{\beta
(\epsilon -\mu )}{2}\right)\right)\frac{\tau }{\pi }, \\
\sigma _{z z}^{\text{dc}}&=&\frac{e^2m v_z}{16\hbar ^3}\frac{\beta }{4}\int \epsilon d\epsilon \left(\text{sech}^2\left(\frac{\beta (\epsilon +\mu )}{2}\right)+\text{sech}^2\left(\frac{\beta
(\epsilon -\mu )}{2}\right)\right)\frac{\tau }{\pi }.
\end{eqnarray}
The dc conductivities show different behaviors in the quantum degenerate limit $\frac{\mu}{k_B T}\gg1$,
\begin{eqnarray}
\label{eq:sigmadc}
\sigma _{x x}^{\text{dc}}=\sigma _{y y}^{\text{dc}}&\approx& \frac{e^2}{3\pi  v_z\hbar ^3}\frac{\tau }{\pi }\left(\mu ^2+\frac{\pi ^2}{3}\left(k_BT\right)^2\right), \\
\sigma _{z z}^{\text{dc}}&\approx& \frac{e^2m v_z}{16\hbar ^3}|\mu |\frac{\tau }{\pi }.
\end{eqnarray}
\end{widetext}
In the $\hat{x}$-$\hat{y}$ plane, the longitudinal electrical conductivity is characterized by a quadratic dependence on both the temperature and the chemical potential; In the $\hat{z}$ direction, the dc conductivity has a Drude peak proportional to the chemical potential. The thermoelectric coefficients are obtained in a similar fashion,
\begin{widetext}
\begin{eqnarray}
L_{x x}^{21}&=&\frac{e}{3\pi  v_z\hbar ^3}\frac{\beta }{4}\int d\epsilon  \left(\text{sech}^2\left(\frac{\beta (\epsilon +\mu )}{2}\right)-\text{sech}^2\left(\frac{\beta
(\epsilon -\mu )}{2}\right)\right)\epsilon ^3\frac{\tau }{\pi } \nonumber \\
&&+\mu \frac{e}{3\pi  v_z\hbar ^3}\frac{\beta }{4}\int d\epsilon  \left(\text{sech}^2\left(\frac{\beta
(\epsilon +\mu )}{2}\right)+\text{sech}^2\left(\frac{\beta (\epsilon -\mu )}{2}\right)\right)\epsilon ^2\frac{\tau }{\pi },\\
L_{z z}^{21}&=&\frac{e m v_z}{16\hbar ^3}\frac{\beta }{4}\int d\epsilon  \left(\text{sech}^2\left(\frac{\beta (\epsilon +\mu )}{2}\right)-\text{sech}^2\left(\frac{\beta
(\epsilon -\mu )}{2}\right)\right)\epsilon ^2\frac{\tau }{\pi } \nonumber \\
&&+\mu \frac{e m v_z}{16\hbar ^3}\frac{\beta }{4}\int \epsilon d\epsilon  \left(\text{sech}^2\left(\frac{\beta
(\epsilon +\mu )}{2}\right)+\text{sech}^2\left(\frac{\beta (\epsilon -\mu )}{2}\right)\right)\frac{\tau }{\pi }.
\end{eqnarray}
At low temperature $\frac{\mu}{k_B T}\gg1$,
\begin{eqnarray}
\label{eq:L21}
L_{x x}^{21} &\approx& -\mu \frac{e}{3\pi  v_z\hbar ^3}\frac{\tau }{\pi }\frac{2\pi ^2}{3}\left(k_BT\right)^2, \nonumber \\
L_{z z}^{21} &\approx& -\frac{e
m v_z}{16\hbar ^3}\frac{\tau }{\pi }\text{sgn}(\mu )\frac{\pi ^2}{3}\left(k_BT\right)^2.
\end{eqnarray}
By comparing Eq.(\ref{eq:L21}) with Eq.(\ref{eq:sigmadc}), the Mott relation $L_{\alpha \alpha}^{21}=-\frac{\pi^2}{3 e}(k_B T)^2 \frac{\partial \sigma _{\alpha \alpha}^{\text{dc}}}{\partial \mu}$ is recovered.\cite{Kargarian:prb13} For $L^{22}$, we have
\begin{eqnarray}
L_{x x}^{22}&=&\frac{1}{3\pi  T \hbar ^3v_z}\frac{\beta }{4}\int d\epsilon  \left[\text{sech}^2\left(\frac{\beta (\epsilon +\mu )}{2}\right)(\epsilon +\mu )^2\epsilon
^2+\text{sech}^2\left(\frac{\beta (\epsilon -\mu )}{2}\right)(\epsilon -\mu )^2\epsilon ^2\right]\frac{\tau }{\pi }, \\
L_{z z}^{22}&=&\frac{m v_z}{16T \hbar ^3}\frac{\beta }{4}\int \epsilon d\epsilon \left[\text{sech}^2\left(\frac{\beta (\epsilon +\mu )}{2}\right)(\epsilon +\mu
)^2+\text{sech}^2\left(\frac{\beta (\epsilon -\mu )}{2}\right)(\epsilon -\mu )^2\right]\frac{\tau }{\pi }.
\end{eqnarray}
\end{widetext}
At low temperature $\frac{\mu}{k_B T}\gg1$,
\begin{eqnarray}
\label{eq:L22}
L_{x x}^{22}&\approx& \frac{1}{3\pi  T \hbar ^3v_z}\frac{\tau }{\pi }\left[\mu ^2 \frac{\pi ^2}{3}\left(k_BT\right)^2+\frac{7\pi ^2}{15}\left(k_BT\right)^4\right],\\
L_{zz}^{22}&\approx& \frac{m v_z}{16T \hbar ^3}\frac{\tau }{\pi }\frac{\pi ^2\left(k_BT\right)^2}{3}|\mu|.
\end{eqnarray}
By comparing Eq.(\ref{eq:L22}) with Eq.(\ref{eq:sigmadc}), one sees that the Wiedemann-Franz law, $L_{\alpha \alpha}^{22}=\frac{\pi^2 k_B^2 T}{3 e^2}\sigma _{\alpha \alpha}^{\text{dc}}$, is observed up to leading order in $k_B T$. From Eq.(\ref{eq:kappa}), one obtains the thermal conductivity as,
\begin{eqnarray}
\kappa _{x x}=\kappa _{y y}&\approx& \frac{\mu^2 k_B^2 T \tau}{9 v_z \hbar^3}(1+\frac{\pi^2 k_B^2 T^2}{15 \mu^2}) ,\nonumber \\
\kappa _{z z}&\approx& \frac{\pi m v_z |\mu| \tau k_B^2 T}{48 \hbar^3}.
\end{eqnarray}
Combining Eq.(\ref{eq:sigmadc}) and Eq.(\ref{eq:L21}), the thermopower
\begin{eqnarray}
S_y=S_x&=&\frac{L_{x x}^{21}}{T L_{x x}^{11}}\approx
-\frac{2\pi ^2k_B^2 T}{3 e \mu }, \nonumber \\ 
S_z&=&\frac{L_{z z}^{21}}{T L_{z z}^{11}}\approx -\frac{\pi ^2k_B^2 T}{3 e \mu },
\end{eqnarray}
shows linear dependence on temperature for $\frac{\mu}{k_B T}\gg1$.

\subsubsection{Anomalous transport properties}

The anomalous thermoelectric transport properties are given in Eq.(\ref{eq:L_anomalous}) with the Berry curvature in the vicinity of double Weyl points evaluated as,
\begin{equation}
\Omega _z^{\chi}=\chi \frac{2 q_{\perp}^2 v_zq_z}{m^2 \epsilon ^3},\\
\\
\Omega _x^{\chi}=\chi \frac{q_{\perp}^2q_xv_z}{m^2 \epsilon ^3},\\
\\
\Omega _y^{\chi}=\chi \frac{q_{\perp}^2q_y v_z}{m^2 \epsilon ^3},
\end{equation}
where $\mathbf{q}$ is the momentum in the vicinity of the Weyl nodes, and $q_{\perp}$ is the component in the $x$-$y$ plane. The anomalous electrical conductivity of a time-reversal symmetry (TRS)  broken realization of double Weyl semimetals when the Fermi energy is at the Weyl nodes is
\begin{equation}
\label{eq: sigmaanomalous}
\sigma _{i j}^A=2\epsilon _{i j l}\frac{e^2}{h}\frac{\Delta  k_l}{\pi},
\end{equation}
where $\Delta  k_l$ is the separation of double-Weyl nodes with opposite chirality in the BZ similar to the case of linear dispersion\cite{Rexprb2014}. The factor of $2$ in front is due to the topological charge of the double-Weyl point. According to the Onsager and Mott relation,\cite{Mermin1976}
\begin{equation}
\label{eq: L21anomalous}
L_{i j}^{21}=T L_{i j}^{12}=\frac{-\pi ^2}{3e}\left(k_BT\right)^2\left(\frac{\partial \sigma _{i j}^A}{\partial \mu }\right)=0.
\end{equation}
The Wiedemann-Franz law also holds for anomalous transport at low temperature,
\begin{equation}
\label{eq: L22anomalous}
L_{i j}^{22}=\frac{\pi ^2}{3}\frac{k_B^2T}{e^2}\sigma _{i j}^A=\frac{2\pi }{3h}k_B^2T\epsilon _{i j l}\Delta  k_l.
\end{equation}
Notice that the anomalous thermoelectric coefficients are proportional to topological charges and do not depend on the details of band dispersion when the Fermi energy is at the double-Weyl point. These transport quantities are dissipationless and Eq.(\ref{eq: sigmaanomalous})-Eq.(\ref{eq: L22anomalous}) also apply to interacting systems\cite{Haldaneprl2004}.

\subsection{Electron-electron interactions}
The electron-electron interaction effect is significant near the charge neutral point $\mu=0$ due to the non-Fermi liquid behavior for a point-like Fermi ``surface"\cite{DasSarmarmp2011}. The model Hamiltonian can be separated as
\begin{equation}
H=H_0+H_1,
\end{equation}
with the free part given by, 
\begin{widetext}
\begin{equation}
H_0=\int \frac{d^3\mathbf{k}}{(2\pi )^3} \Psi ^{\dagger }(\mathbf{k})\left(\frac{k_x^2-k_y^2}{m}\sigma _x+\frac{2k_xk_y}{m}\sigma _y+\chi v_zk_z\sigma
_z\right)\Psi (\mathbf{k}),
\end{equation}
and the electron-electron interaction term,
\begin{equation}
H_1=\frac{1}{2}\int \frac{d^3\mathbf{k}_1}{(2\pi )^3}\frac{d^3\mathbf{k}_2}{(2\pi )^3}\frac{d^3\mathbf{q}}{(2\pi )^3}\Psi ^{\dagger }\left(\mathbf{k}_2-\mathbf{q}\right)\Psi
\left(\mathbf{k}_2\right)V(\mathbf{q})\Psi ^{\dagger }\left(\mathbf{k}_1+\mathbf{q}\right)\Psi \left(\mathbf{k}_1\right).
\end{equation}
\end{widetext}
The anisotropically screened Coulomb interaction at long wavelength is,\cite{Hsin-Huaprb2015, Shao-Kaiprb2015}
\begin{equation}
\label{eq:Vee}
V(\mathbf{q})=\frac{4\pi  e^2}{\epsilon \left(q_{\perp}^2+\eta \left|q_z\right|\right)},
\end{equation}
where $\eta $ is the anisotropic factor. At low temperature, due to the critical nature of the system\cite{Fritzprb2008, Burkovprb2011}, the relaxation time can be obtained from dimensional analysis similar to the case of linear dispersion\cite{Hosurprl2012, Burkovprb2011}. Notice that in Eq.(\ref{eq:Vee}), $q_{\perp}^2$ and $\left|q_z\right|$ have the same scaling dimension. As a result, we have
\begin{equation}
\label{eq: tauee}
\tau _{\perp}=\frac{A_{\perp}}{k_BT}, \\
\tau _z=\frac{A_z}{k_BT},
\end{equation}
up to logarithmic corrections,\cite{Hsin-Huaprb2015} for the transverse and longitudinal relaxation times. The constants $A_z$ and $A_{\perp}$ have a linear dependence on $(\frac{e^2}{4\pi \epsilon})^2$. For $\frac{\mu}{k_B T}\ll1$, the longitudinal electrical conductivity is obtained as
\begin{eqnarray}
\label{eq:sigmadcee}
\sigma _{x x}^{\text{dc}}=\sigma _{y y}^{\text{dc}}&\approx& \frac{\pi e^2 (k_B T)^2}{9 v_z \hbar^3} \frac{\tau_{\perp}}{\pi} (1+\frac{3}{\pi^2}(\frac{\mu}{k_B T})^2) \nonumber \\
&=&\frac{A_{\perp} e^2 k_B T}{9 v_z \hbar^3}(1+\frac{3}{\pi^2}(\frac{\mu}{k_B T})^2), \nonumber \\
\sigma _{z z}^{\text{dc}}&\approx& \frac{e^2 m v_z}{8 \hbar^3}\frac{\tau_z}{\pi} (k_B T) (\mathrm{ln(2)}+\frac{1}{8}(\frac{\mu}{k_B T})^2) \nonumber \\
&=&\frac{A_z e^2 m v_z}{8 \pi \hbar^3}(\mathrm{ln(2)}+\frac{1}{8}(\frac{\mu}{k_B T})^2), 
\end{eqnarray}
of which the temperature dependence is consistent with the RG analysis in Ref.[\onlinecite{Hsin-Huaprb2015}] up to logarithmic corrections in temperature. Likewise, other thermoelectric coefficients can be obtained by replacing the relaxation time in the free electron case with Eq.(\ref{eq: tauee}),
\begin{eqnarray}
\label{eq:L21ee}
L_{x x}^{21} =L_{y y}^{21}&\approx& -\mu \frac{2\pi e}{9 v_z \hbar^3} \frac{\tau_{\perp}}{\pi} (k_B T)^2 \nonumber \\
&=& -\mu \frac{2 A_{\perp} e}{9 v_z \hbar^3} k_B T, \nonumber \\
L_{z z}^{21} &\approx& -\mu \frac{e m v_z}{8 \hbar^3} \mathrm{ln(2)} \frac{\tau_z}{\pi} (k_B T) \nonumber \\
&=& -\mu \frac{A_{z} e m v_z}{8 \pi \hbar^3} \mathrm{ln(2)},
\end{eqnarray}
\begin{eqnarray}
\label{eq:L22ee}
L_{x x}^{22}=L_{y y}^{22}&\approx& \frac{1}{3\pi T \hbar ^3v_z} \frac{\tau_{\perp}}{\pi} \left(\frac{7 \pi ^4}{15}\left(k_BT\right)^4+\mu ^2\frac{\pi ^2}{3}(k_BT)^2\right) \nonumber \\
&=&\frac{A_{\perp}}{3\pi ^2 T \hbar ^3v_z}\left(\frac{7 \pi ^4}{15}\left(k_BT\right)^3+\mu ^2\frac{\pi ^2}{3}k_BT\right) ,\nonumber \\
L_{z z}^{22}&\approx&\frac{9 \zeta(3) m v_z}{16 T \hbar ^3} \frac{\tau_{z}}{\pi} \left(k_BT\right)^3 \nonumber \\
&=&\frac{9 \zeta(3) A_z m v_z}{16\pi  T \hbar^3}\left(k_BT\right)^2.
\end{eqnarray}
From Eq.(\ref{eq:kappa}), one can obtain the thermal conductivity as,
\begin{eqnarray}
\label{eq:kappaee}
\kappa _{x x}=\kappa _{y y}&\approx& \frac{\pi (k_B T)^4}{45 v_z \hbar^3 T} \frac{\tau_{\perp}}{\pi} (7\pi^2-15(\frac{\mu}{k_B T})^2) \nonumber \\
&=& \frac{A_{\perp} (k_B T)^3}{45 v_z \hbar^3 T} (7\pi^2-15(\frac{\mu}{k_B T})^2), \nonumber \\
\kappa _{z z}&\approx& \frac{m v_z}{16 T \hbar^3} (k_B T)^3 \frac{\tau_z}{\pi} (9 \zeta(3)-2 \mathrm{ln(2)} (\frac{\mu}{k_B T})^2) \nonumber \\
&=& \frac{A_z m v_z}{16 \pi T \hbar^3} (k_B T)^2 (9 \zeta(3)-2 \mathrm{ln(2)} (\frac{\mu}{k_B T})^2).
\end{eqnarray}
Combining Eq.(\ref{eq:sigmadcee}) and Eq.(\ref{eq:L21ee}), the thermopower
\begin{equation}
\label{eq:Seebeckee}
S_x=S_y \approx 2 S_z \approx -\frac{2 \mu}{e T}.
\end{equation}
Our results for the longitudinal transport quantities in $x$-$y$ plane from Eq.(\ref{eq:sigmadcee}) to Eq.(\ref{eq:kappaee}) double that of single Weyl node with linear dispersion,\cite{Rexprb2014} while the thermopower $S_x, S_y$ and $S_z$ are still independent of model parameters at $\frac{\mu}{k_B T}\ll1$.  We emphasize that the results above assume negligible inter-node scattering, and therefore neglect the physics of the chiral anomaly.  We will discuss the case of relevant inter-node scattering in Sec. III E.

\subsection{Diffusive transport}
By doping away from the double-Weyl node, disorder effects become more important. Here we assume a model with short-range impurity scattering potential of the form
\begin{equation}
V(\mathbf{r})=u_0\sum _a \delta \left(\mathbf{r}-\mathbf{r}_a\right),
\end{equation}
where $u_0$ is the disorder potential strength and $\mathbf{r}_a$ labels the random impurity positions.  Here we neglect electron-electron interactions. By considering the first Born approximation for electron lifetime under impurity scattering,
\begin{eqnarray}
\frac{1}{\tau _s(\mathbf{k})}&=&\frac{2\pi  n_i}{\hbar }\int \frac{d^3\mathbf{k}'}{(2\pi )^3}\delta \left(\epsilon_{s\bf k}-\epsilon_{s\bf k'}\right)|T_{\mathbf{k}\mathbf{
}\mathbf{k}'}|^2 \nonumber \\
&=&\frac{\pi }{2\hbar }u_0^2n_ig(\epsilon ),
\end{eqnarray}
where $T_{\mathbf{k} \mathbf{k}'}=u_0 \langle s,\mathbf{k}|s, \mathbf{k}' \rangle$ is the scattering matrix element. The transport lifetime differs from the electron lifetime and should be anisotropic due to the anisotropy of the Fermi surface\cite{Adroguer2015}. By examining the Boltzmann equation,
\begin{equation}
\label{eq:Boltamanndiffusive}
\frac{-e}{\hbar }\mathbf{E}\mathbf{\cdot }\mathbf{\nabla }_{\mathbf{k}}f=\frac{2\pi  n_i}{\hbar }\int \frac{d^3\mathbf{k}'}{(2\pi )^3}\delta \left(\epsilon
_{\mathbf{k}}-\epsilon _{\mathbf{k}'}\right)|T_{\mathbf{k}\mathbf{ }\mathbf{k}'}|^2\left(f\left(\mathbf{k}'\right)-f(\mathbf{k})\right),
\end{equation}
with
\begin{equation}
f(\mathbf{k}) \approx n_F\left(\epsilon _{\mathbf{k}}\right)+e\frac{\partial n_F}{\partial
\epsilon }\mathbf{\Lambda }(\mathbf{k})\cdot \mathbf{E},
\end{equation}
we have
\begin{eqnarray}
\mathbf{v}_{\mathbf{k}}&=&\frac{1}{\tau (\mathbf{k})}\mathbf{\Lambda }(\mathbf{k}) \nonumber \\
&&-\frac{ 2\pi  n_iu_0^2}{\hbar }\int \frac{d^3\mathbf{k}'}{(2\pi )^3}\delta \left(\epsilon
_{\mathbf{k}}-\epsilon _{\mathbf{k}'}\right) \times \nonumber \\
&&|\left\langle \epsilon (\mathbf{k})\left|\epsilon \left(\mathbf{k}'\right)\right.\right\rangle |^2\mathbf{\Lambda }\left(\mathbf{k}'\right).
\end{eqnarray}
The transport lifetime is defined as,
\begin{equation}
\Lambda _{\alpha }\left(\mathbf{k}'\right)=v_{\alpha }\left(\mathbf{k}'\right)\tau _{\alpha }^{\text{tr}}\left(\mathbf{k}'\right).
\end{equation}
By making the ansatz,
\begin{equation}
\label{eq: tauansatz}
\tau _{\alpha }^{\text{tr}}=\lambda _{\alpha }(\epsilon )\tau (\mathbf{k}),
\end{equation}
and plugging Eq.(\ref{eq: tauansatz}) into Eq.(\ref{eq:Boltamanndiffusive}), we could determine $\lambda _{\alpha }(\epsilon)$ as 
\begin{eqnarray}
\lambda _x(\epsilon )&=&\lambda _y(\epsilon )=1, \nonumber \\
\lambda _z(\epsilon )&=&2,
\end{eqnarray}
which gives
\begin{equation}
\label{eq:tautransport}
\tau _x^{\text{tr}}(\epsilon )=\tau _y^{\text{tr}}(\epsilon )=\frac{\tau _z^{\text{tr}}(\epsilon )}{2}=\tau (\epsilon ).
\end{equation}
We use Eq.(\ref{eq:tautransport}) for the relaxation time to calculate thermoelectric quantities
\begin{widetext}
\begin{eqnarray}
\sigma _{x x}^{\text{dc}}=\sigma _{y y}^{\text{dc}}&=&e^2\sum _s \int \frac{d^3\mathbf{k}}{(2\pi )^3} \tau _x^{\text{tr}}\left(\epsilon _{s\bf k})\right)\left(-\frac{\partial
f}{\partial \epsilon _s}\right)\left(\frac{1}{\hbar }\frac{\partial \epsilon_{s\bf k})}{\partial k_x}\right)^2 \nonumber \\
&=&\frac{8v_z\hbar ^4}{u_0^2m n_i}\frac{m e^2}{3\pi ^2m v_z\hbar ^3}\frac{\beta }{4}\int d\epsilon  \left(\text{sech}^2\left(\frac{\beta (\epsilon
+\mu )}{2}\right)+\text{sech}^2\left(\frac{\beta (\epsilon -\mu )}{2}\right)\right)\epsilon, \\
\sigma _{z z}^{\text{dc}}&=&e^2\sum _s \int \frac{d^3\mathbf{k}}{(2\pi )^3} \tau _z^{\text{tr}}\left(\epsilon _{s\bf k}\right)\left(-\frac{\partial
f}{\partial \epsilon _s}\right)\left(\frac{1}{\hbar }\frac{\partial \epsilon_{s\bf k}}{\partial k_z}\right)^2 \nonumber \\
&=&\frac{16v_z\hbar ^4}{u_0^2m n_i}\frac{e^2m v_z}{16\pi  \hbar ^3}\frac{\beta }{4}\int d\epsilon \left(\text{sech}^2\left(\frac{\beta (\epsilon
+\mu )}{2}\right)+\text{sech}^2\left(\frac{\beta (\epsilon -\mu )}{2}\right)\right).
\end{eqnarray}
For $\frac{\mu}{k_B T}\ll1$,
\begin{eqnarray}
\sigma _{x x}^{\text{dc}}=\sigma _{y y}^{\text{dc}}&\approx&\frac{8 e^2 \hbar}{3 \pi ^2 m n_i u_0^2} k_BT (2\mathrm{ln(2)}+\frac{1}{4}(\frac{\mu}{k_B T})^2), \nonumber \\
\sigma _{z z}^{\text{dc}}&\approx&\frac{e^2 v_z^2 \hbar}{\pi n_i u_0^2}.
\end{eqnarray}
For $\frac{\mu}{k_B T}\gg1$,
\begin{eqnarray}
\label{eq:sigmadisorder}
\sigma _{x x}^{\text{dc}}=\sigma _{y y}^{\text{dc}}&\approx&\frac{8 e^2 \hbar}{3 \pi ^2 m n_i u_0^2} |\mu|, \nonumber \\
\sigma _{z z}^{\text{dc}}&\approx&\frac{e^2 v_z^2 \hbar}{\pi n_i u_0^2}.
\end{eqnarray}
\end{widetext}
In Appendix \ref{sec: diffusive diagram}, we calculated the longitudinal electrical conductivity by a diagrammatic approach including vertex corrections to current operators. Up to leading order in disorder strength $u_0^2 n_i$, the results are in agreement with Eq.(\ref{eq:sigmadisorder}). 

The other thermoelectric coefficients can be obtained in the same fashion,
\begin{widetext}
\begin{eqnarray}
L_{x x}^{21}=L_{y y}^{21}&=&-e\sum _s \int \frac{d^3\mathbf{k}}{(2\pi )^3}\text{  }\tau _x^{\text{tr}}\left(\epsilon_{s\bf k}\right)\left(-\frac{\partial f}{\partial \epsilon
_s}\right)\left(\epsilon_{s\bf k}-\mu \right)\left(\frac{1}{\hbar }\frac{\partial \epsilon_{s\bf k}}{\partial k_x}\right)^2 \nonumber \\
&=&\frac{8v_z\hbar ^4}{u_0^2m n_i}\frac{e}{3\pi ^2 v_z\hbar ^3}\frac{\beta }{4}\int d\epsilon  \left(\text{sech}^2\left(\frac{\beta (\epsilon +\mu
)}{2}\right)-\text{sech}^2\left(\frac{\beta (\epsilon -\mu )}{2}\right)\right)\epsilon ^2 \nonumber \\
&&+\mu \frac{8v_z\hbar ^4}{u_0^2m n_i}\frac{e}{3\pi ^2 v_z\hbar
^3}\frac{\beta }{4}\int d\epsilon  \left(\text{sech}^2\left(\frac{\beta (\epsilon +\mu )}{2}\right)+\text{sech}^2\left(\frac{\beta (\epsilon -\mu
)}{2}\right)\right)\epsilon, \\
L_{z z}^{21}&=&-e\sum _s \int \frac{d^3\mathbf{k}}{(2\pi )^3}\text{  }\tau _z^{\text{tr}}\left(\epsilon_{s\bf k}\right)\left(-\frac{\partial f}{\partial \epsilon_s}\right)\left(\epsilon_{s\bf k}-\mu \right)\left(\frac{1}{\hbar }\frac{\partial \epsilon_{s\bf k}}{\partial k_z}\right)^2 \nonumber \\
&=&\frac{8v_z\hbar ^4}{u_0^2m n_i}\frac{2e m v_z}{16\pi  \hbar ^3}\frac{\beta }{4}\int d\epsilon  \left(\text{sech}^2\left(\frac{\beta (\epsilon+\mu )}{2}\right)-\text{sech}^2\left(\frac{\beta (\epsilon -\mu )}{2}\right)\right)\epsilon \nonumber \\
&&+\mu \frac{8v_z\hbar ^4}{u_0^2m n_i}\frac{2e m v_z}{16\pi\hbar ^3}\frac{\beta }{4}\int d\epsilon \left(\text{sech}^2\left(\frac{\beta (\epsilon +\mu )}{2}\right)+\text{sech}^2\left(\frac{\beta (\epsilon-\mu )}{2}\right)\right), 
\end{eqnarray}
\begin{eqnarray}
L_{x x}^{22}=L_{yy}^{22}&=&\frac{1}{T}\sum _s \int \frac{d^3\mathbf{k}}{(2\pi )^3} \tau _x^{\text{tr}}\left(\epsilon_{s\bf k}\right)\left(-\frac{\partial f}{\partial \epsilon
_s}\right)\left(\epsilon_{s\bf k}-\mu \right)^2\left(\frac{1}{\hbar }\frac{\partial \epsilon_{s\bf k}}{\partial k_x}\right)^2 \nonumber \\
&=&\frac{8v_z\hbar ^4}{u_0^2m n_i}\frac{1}{3\pi ^2 T \hbar ^3v_z}\frac{\beta }{4}\int d\epsilon  \epsilon \left[\text{sech}^2\left(\frac{\beta (\epsilon
+\mu )}{2}\right)(\epsilon +\mu )^2+\text{sech}^2\left(\frac{\beta (\epsilon -\mu )}{2}\right)(\epsilon -\mu )^2\right], \\
L_{z z}^{22}&=&\frac{1}{T}\sum _s \int \frac{d^3\mathbf{k}}{(2\pi )^3} \tau _z^{\text{tr}}\left(\epsilon_{s\bf k}\right)\left(-\frac{\partial f}{\partial \epsilon
_s}\right)\left(\epsilon_{s\bf k}-\mu \right)^2\left(\frac{1}{\hbar }\frac{\partial \epsilon_{s\bf k}}{\partial k_z}\right)^2 \nonumber \\
&=&\frac{16v_z\hbar ^4}{u_0^2m n_i}\frac{m v_z}{16 \pi  T \hbar ^3}\frac{\beta }{4}\int d\epsilon \left[\text{sech}^2\left(\frac{\beta (\epsilon
+\mu )}{2}\right)(\epsilon +\mu )^2+\text{sech}^2\left(\frac{\beta (\epsilon -\mu )}{2}\right)(\epsilon -\mu )^2\right].
\end{eqnarray}
\end{widetext}
For $\frac{\mu}{k_B T}\ll1$,
\begin{eqnarray}
L_{x x}^{21}=L_{y y}^{21}&\approx&-\mu \frac{8 e \hbar }{3\pi ^2 n_i u_0^2 m} 2\mathrm{ln(2)}k_BT, \nonumber \\
L_{z z}^{21}&\approx& 0, \nonumber \\
L_{x x}^{22}=L_{y y}^{22}&\approx& \frac{8 \hbar}{\pi ^2 n_i u_0^2 m T} 3 \zeta(3)\left(k_BT\right)^3 ,\nonumber \\
L_{z z}^{22}&\approx& \frac{v_z^2\hbar}{\pi n_i u_0^2 T}\frac{\pi ^2}{3}\left(k_BT\right)^2.
\end{eqnarray}
For $\frac{\mu}{k_B T}\gg1$,
\begin{eqnarray}
\label{eq:L21disorder}
L_{x x}^{21}=L_{y y}^{21}&\approx& -\frac{8 e \hbar }{3\pi ^2 n_i u_0^2 m} \frac{\pi ^2}{3}\text{sgn}(\mu )\left(k_BT\right)^2 ,\nonumber \\
L_{z z}^{21}&\approx& 0, \nonumber \\
L_{x x}^{22}=L_{y y}^{22}&\approx& \frac{8 \hbar}{\pi ^2 n_i u_0^2 m T} |\mu | \frac{\pi ^2}{3}\left(k_BT\right)^2, \nonumber \\
L_{z z}^{22}&\approx& \frac{v_z^2\hbar}{\pi n_i u_0^2 T}\frac{\pi ^2}{3}\left(k_BT\right)^2.
\end{eqnarray}
From Eq.(\ref{eq:kappa}), one can obtain the thermal conductivity for $\frac{\mu}{k_B T}\ll1$,
\begin{eqnarray}
\kappa _{x x}=\kappa _{yy}&\approx& \frac{8 \hbar k_B^3 T^2}{\pi^2 n_i u_0^2 m} (3\zeta(3)-\frac{2 \mathrm{ln(2)}}{3}(\frac{\mu}{k_B T})^2),\nonumber \\
\kappa _{z z}&\approx& \frac{v_z^2\hbar}{\pi n_i u_0^2 T}\frac{\pi ^2}{3}\left(k_BT\right)^2.
\end{eqnarray}
and for $\frac{\mu}{k_B T}\gg1$,
\begin{eqnarray}
\kappa _{x x}=\kappa _{yy}&\approx& \frac{8 \hbar k_B^2 T}{9 n_i u_0^2 m}|\mu|(1-\frac{\pi^2}{3}(\frac{k_B T}{\mu})^2), \nonumber \\
\kappa _{z z}&\approx& \frac{v_z^2\hbar}{\pi n_i u_0^2 T}\frac{\pi ^2}{3}\left(k_BT\right)^2.
\end{eqnarray}
Combining Eq.(\ref{eq:sigmadisorder}) and Eq.(\ref{eq:L21disorder}), the thermopower is calculated as,
\begin{eqnarray}
S_x=S_y&=&\frac{L_{x x}^{21}}{T L_{x x}^{11}}\approx -\frac{\pi^2 k_B^2 T}{3 e \mu}, \nonumber \\ 
S_z&=&\frac{L_{z z}^{21}}{T L_{z z}^{11}}\approx 0.
\end{eqnarray}
Notice that $S_z\approx0$ due to $L^{21}_{z z}=0$ is an artifact in first Born approximation. A more physical estimate of $L^{21}_{z z}$ and $S_{z}$ might be obtained by the self-consistent Born approximation\cite{Rexprb2014}.

\begin{table*}[h!]
\begin{center}
\begin{tabular*}{\textwidth}{| P{0.161\textwidth} | P{0.045\textwidth} | P{0.3\textwidth} | P{0.35\textwidth} | P{0.1\textwidth} |}
\hline
&$\alpha \alpha$ & $\sigma_{\alpha \alpha}$ & $\kappa_{\alpha \alpha}$ & $S_\alpha$ \\
\hline
Free electron &$x x$&$\frac{e^2}{3\pi  v_z\hbar ^3}\frac{\tau }{\pi }\left(\mu ^2+\frac{\pi ^2}{3}\left(k_BT\right)^2\right)$ & $\frac{\mu^2 k_B^2 T \tau}{9 v_z \hbar^3}(1+\frac{\pi^2 k_B^2 T^2}{15 \mu^2})$ & $-\frac{2\pi ^2k_B^2 T}{3 e \mu }$\\
\cline{2-5}
$\frac{\mu}{k_B T}\gg1$&$z z$&$\frac{e^2m v_z}{16\hbar ^3}|\mu |\frac{\tau }{\pi }$& $\frac{\pi m v_z |\mu| \tau k_B^2 T}{48 \hbar^3}$ & $-\frac{\pi ^2k_B^2 T}{3 e \mu }$\\
\cline{2-5}
&$x x$&$\frac{ \pi  e^2}{9 v_z\hbar ^3}\left(k_BT\right)^2\frac{\tau }{\pi }\left(1+\frac{3}{\pi ^2}\left(\frac{\mu }{k_BT}\right)^2\right)$&$\frac{\tau \left(k_B T\right)^4}{45 v_z \hbar ^3T}\left(7\pi ^2-15\left(\frac{\mu }{k_BT}\right)^2\right)$&$-\frac{2 \mu}{e T}$\\
\cline{2-5}
$\frac{\mu}{k_B T}\ll1$&$z z$&$\frac{e^2m v_z}{8\hbar ^3}k_BT\frac{\tau }{\pi }\left(\mathrm{ln(2)}+\frac{1}{8}\left(\frac{\mu }{k_BT}\right)^2\right)$&$\frac{m v_z \tau}{16 \pi T \hbar ^3}\left(k_BT\right)^3 \left(9 \zeta(3)-2\mathrm{ln(2)}\left(\frac{\mu }{k_BT}\right)^2\right)$&$-\frac{\mu}{e T}$\\
\hline
e-e interaction &$x x$&$\frac{A_{\perp} e^2 k_B T}{9 v_z \hbar^3}(1+\frac{3}{\pi^2}(\frac{\mu}{k_B T})^2)$&$\frac{A_{\perp} (k_B T)^3}{45 v_z \hbar^3 T} (7\pi^2-15(\frac{\mu}{k_B T})^2)$&$-\frac{2 \mu}{e T}$\\
\cline{2-5}
$\frac{\mu}{k_B T}\ll1$&$z z$&$\frac{A_z e^2 m v_z}{8 \pi \hbar^3}(\mathrm{ln(2)}+\frac{1}{8}(\frac{\mu}{k_B T})^2)$&$\frac{A_z m v_z}{16 \pi T \hbar^3} (k_B T)^2 (9 \zeta(3)-2 \mathrm{ln(2)} (\frac{\mu}{k_B T})^2)$&$-\frac{\mu}{e T}$ \\
\hline
Diffusion &$x x$&$\frac{8 e^2 \hbar}{3 \pi ^2 m n_i u_0^2} |\mu|$&$\frac{8 \hbar k_B^2 T}{9 n_i u_0^2 m}|\mu|(1-\frac{\pi^2}{3}(\frac{k_B T}{\mu})^2)$&$-\frac{\pi^2 k_B^2 T}{3 e \mu}$\\
\cline{2-5}
$\frac{\mu}{k_B T}\gg1$&$z z$&$\frac{e^2 v_z^2 \hbar}{\pi n_i u_0^2}$&$\frac{v_z^2\hbar}{\pi n_i u_0^2 T}\frac{\pi ^2}{3}\left(k_BT\right)^2$& $0$\\
\cline{2-5}
&$x x$&$\frac{8 e^2 \hbar}{3 \pi ^2 m n_i u_0^2} k_BT (2\mathrm{ln(2)}+\frac{1}{4}(\frac{\mu}{k_B T})^2)$&$\frac{8 \hbar k_B^3 T^2}{\pi^2 n_i u_0^2 m} (3\zeta(3)-\frac{2 \mathrm{ln(2)}}{3}(\frac{\mu}{k_B T})^2)$&$-\frac{\mu}{e T}$\\
\cline{2-5}
$\frac{\mu}{k_B T}\ll1$&$z z$&$\frac{e^2 v_z^2 \hbar}{\pi n_i u_0^2}$&$\frac{v_z^2\hbar}{\pi n_i u_0^2 T}\frac{\pi ^2}{3}\left(k_BT\right)^2$&$0$ \\
\hline
\end{tabular*}
\end{center}
\caption{Transport coefficients for different collision mechanism in the absence of a magnetic field.  Here $\sigma_{\alpha \alpha}$ is the electrical conductivity, $\kappa_{\alpha \alpha}$ the thermal conductivity, and $S_\alpha$ the Seebeck coefficient, with spatial components labeled by $\alpha$.}
\label{table:nofield}
\end{table*}

\subsection{Discussion on transport with no field}

Before we move on to the thermoelectric properties under a magnetic field, we briefly summarize and discuss the results obtained for the case of zero field.  They are listed in Table~\ref{table:nofield}.  By comparing the diffusive thermoelectric transport coefficients with those under electron-electron interaction, one sees that in the classical limit $\mu/k_BT \ll 1$, they give rise to a similar temperature dependence in the transport quantities, which means both collision processes have similar impacts on the thermoelectric properties at low doping. In real materials, transport will be decided by a mix of all scattering mechanism. If one assumes independent scattering processes, then according to Matthiessen's rule, the total scattering rate is $1/\tau_{total}=\sum_i 1/\tau_i$, with $\tau_i$ the rates from different relaxation processes. For $\mu/k_BT \gg 1$, we expect Fermi liquid behavior to hold for double-Weyl fermions.  In this regime, the quasiparticle is well defined and the thermoelectric transport is dominated by disorder effects because $\tau_{dis}/\tau_{e-e}\sim(\frac{k_B T}{\mu})^2\rightarrow0$. 

We would like to mention that due to the poor screening of charged impurities when the system is doped away from the double-Weyl point, the disorder potential will be both anisotropic and longer-ranged, which might become dominate for a certain energy range. 

\
\
\section{THERMOELECTRIC TRANSPORT UNDER MAGNETIC FIELDS}
\label{sect: with field}

We consider an external magnetic field exerted on two double Weyl points with opposite chirality due to TRS breaking,
\begin{eqnarray}
H^{\chi}&=&\frac{k_x^2-k_y^2}{m}\sigma _x+\frac{2k_xk_y}{m}\sigma _y+\chi v_zk_z\sigma _z,
\end{eqnarray}
with an additional Zeeman term,
\begin{equation}
H_{\text{ext}}=g \mathbf{B\cdot \sigma },
\end{equation}
which will not split the double Weyl point into two single Weyl points when $\mathbf{B}$ is applied in the $\hat{z}$-direction.  If applied
in the x-y plane it will split the double Weyl point into two single Weyl points with the same chirality, and the results of the previous sections that ignore inter-node scattering will apply. Our focus is on \(\mathbf{B}=B \hat{z}\). 
The Boltzman transport equation, Eq.(\ref{eq:Boltzmann}), can be rewritten as
\begin{widetext}
\begin{eqnarray}
\label{eq:Boltzmann2}
f^{\chi }&=&f_{\text{eq}}-\left(1+\frac{e}{c}\mathbf{B}\cdot \mathbf{\Omega}^{\chi}\right)^{-1}\tau \left(v_{\mathbf{p}}+e \mathbf{E}\times \mathbf{\Omega}^{\chi}+\frac{e}{c}\left(\mathbf{v}_{\mathbf{p}}\cdot
\mathbf{\Omega}^{\chi}\right)\mathbf{B}\right)\mathbf{\cdot }\mathbf{\nabla }_{\mathbf{r}}f^{\chi } \nonumber \\
&&+\left(1+\frac{e}{c}\mathbf{B}\cdot \mathbf{\Omega}^{\chi}\right)^{-1}\tau  \left(e
\mathbf{E}+\frac{e}{c}\mathbf{v}_{\mathbf{p}}\times \mathbf{B}+\frac{e^2}{c}(\mathbf{B\cdot E})\mathbf{\Omega}^{\chi}\right)\cdot \nabla _{\mathbf{p}}f^{\chi }.
\end{eqnarray}
\end{widetext}
For simplicity, we assumed an isotropic relaxation time and neglected inter-node scattering. The normal velocity is given by
\begin{equation}
v_{\mathbf{p}}^x=\frac{2 k_x k_{\perp}^2}{m^2 \epsilon }, v_{\mathbf{p}}^y=\frac{2 k_y k_{\perp}^2}{m^2 \epsilon }, v_{\mathbf{p}}^z=\frac{k_z v_z^2}{\epsilon }.
\end{equation}
and the Berry curvature can be approximated as
\begin{equation}
\label{eq: berry}
\Omega _x^{\chi } \approx \chi \frac{k_{\perp}^2\left(v_zk_x\right)}{m^2\epsilon ^3},
\Omega _y^{\chi } \approx \chi  \frac{k_{\perp}^2\left(v_zk_y\right)}{m^2\epsilon^3},
\Omega _z^{\chi } \approx \chi  \frac{2k_{\perp}^2\left(v_zk_z\right)}{m^2\epsilon ^3},
\end{equation}
where we have neglected the shift of the double Weyl node in the presence of a Zeeman term in the weak magnetic field limit. Eq.(\ref{eq:Boltzmann2}) can be solved iteratively in the linear response regime, and the solutions are examined case by case in the following. The magneto-transport results are summarized in Table. \ref{table:withfield}. 

\subsection{$\mathbf{B}=B \hat{z}, \mathbf{E}=E \hat{z}, \mathbf{\nabla} T=0$}
\label{subsec: A}

First we consider the electrical transport when the electrical field is parallel to the magnetic field and no temperature gradient is present. Following Ref.[\onlinecite{KiSeokprb2014}], we make the following ansatz in solving the non-equilibrium distribution function for a given chirality,
\begin{widetext}
\begin{equation}
\label{eq:ansatz1}
f^{\chi }=f_{\text{eq}}+\left(1+\frac{e}{c}\mathbf{B}\cdot \mathbf{\Omega}^{\chi}\right)^{-1}\tau  \left(e E\nabla _{p_z}f_{\text{eq}}+\frac{e^2}{c}E B
\left(\mathbf{\Omega }\cdot \nabla _{\mathbf{p}}\right)f_{\text{eq}}\right)+\left(-\frac{\partial f_{\text{eq}}}{\partial \epsilon }\right)\mathbf{v}\mathbf{\cdot
}\mathbf{\Lambda }^{\chi }.
\end{equation}
with $\mathbf{\Lambda }^{\chi }$ depending on $k_{\perp}$ and $k_z$. By plugging Eq.(\ref{eq:ansatz1}) into Eq.(\ref{eq:Boltzmann2}), one finds
\begin{eqnarray}
\Lambda_x^{\chi }&=&\frac{\tau  \frac{e B}{m c}\tau  \frac{e B}{c} \frac{2k_{\perp}^2}{m \epsilon }\Omega _y^{\chi}+\left(\tau  \frac{e
B}{m c}\frac{2k_{\perp}^2}{m \epsilon }\right)^2\left(1+\frac{e}{c}\mathbf{B}\cdot \mathbf{\Omega }^{\chi}\right)^{-1}\tau  \frac{e B}{c} \Omega
_x^{\chi}}{\left(1+\frac{e}{c}\mathbf{B}\cdot \mathbf{\Omega }^{\chi}\right)^2+\left(\tau  \frac{e B}{m c}\frac{2k_{\perp}^2}{m \epsilon }\right)^2}e E, \nonumber \\
 \Lambda_y^{\chi }&=&\frac{-\tau  \frac{e B}{m c}\tau  \frac{e B}{c} \frac{2k_{\perp}^2}{m \epsilon }\Omega _x^{\chi}+\left(\tau  \frac{e
B}{m c}\frac{2k_{\perp}^2}{m \epsilon }\right)^2\left(1+\frac{e}{c}\mathbf{B}\cdot \mathbf{\Omega }^{\chi}\right)^{-1}\tau  \frac{e B}{c} \Omega
_y^{\chi}}{\left(1+\frac{e}{c}\mathbf{B}\cdot \mathbf{\Omega }^{\chi}\right)^2+\left(\tau  \frac{e B}{m c}\frac{2k_{\perp}^2}{m \epsilon }\right)^2}e E, \nonumber \\
 \Lambda_z^{\chi }&=&0.
\end{eqnarray}
The electrical and thermal currents are given by\cite{KiSeokprb2014, DiXiaormp2010}
\begin{eqnarray}
\mathbf{J}&=&-e\int \frac{d^3p}{(2\pi )^3}\left(1+\frac{e}{c}\mathbf{B}\cdot \mathbf{\Omega }^{\chi}\right)\dot{\mathbf{r}}f^{\chi } \nonumber \\
&=&-e\int \frac{d^3p}{(2\pi )^3}\left(\mathbf{v}_{\mathbf{p}}+e \mathbf{E}\times \mathbf{\Omega }^{\chi}+\frac{e}{c}\left(\mathbf{v}_{\mathbf{p}}\cdot \mathbf{\Omega }^{\chi}\right)\mathbf{B}\right) \times \nonumber \\&&\left(f_{\text{eq}}-\left(1+\frac{e}{c}\mathbf{B}\cdot
\mathbf{\Omega }^{\chi}\right)^{-1}\tau  \left(v_{p_z}+\frac{e}{c} B \left(\mathbf{\Omega }^{\chi}\cdot \mathbf{v}_{\mathbf{p}}\right)\right)\left(-\frac{\partial f_{\text{eq}}}{\partial
\epsilon }\right)e E+\left(-\frac{\partial f_{\text{eq}}}{\partial \epsilon }\right)\mathbf{v}_{\mathbf{p}}\mathbf{\cdot }\mathbf{\Lambda }^{\chi }\right), \\
\mathbf{J}^Q&=&-e\int \frac{d^3p}{(2\pi )^3}(\epsilon -\mu )\left(1+\frac{e}{c}\mathbf{B}\cdot \mathbf{\Omega }^{\chi}\right)\dot{\mathbf{r}}f^{\chi } \nonumber \\
&=&-e\int \frac{d^3p}{(2\pi )^3}(\epsilon -\mu )\left(\mathbf{v}_{\mathbf{p}}+e \mathbf{E}\times \mathbf{\Omega }^{\chi}+\frac{e}{c}\left(\mathbf{v}_{\mathbf{p}}\cdot \mathbf{\Omega }^{\chi}\right)\mathbf{B}\right) \times \nonumber \\
&&\left(f_{\text{eq}}-\left(1+\frac{e}{c}\mathbf{B}\cdot
\mathbf{\Omega }^{\chi}\right)^{-1}\tau  \left(v_{p_z}+\frac{e}{c} B \left(\mathbf{\Omega }^{\chi}\cdot \mathbf{v}_{\mathbf{p}}\right)\right)\left(-\frac{\partial f_{\text{eq}}}{\partial
\epsilon }\right)e E+\left(-\frac{\partial f_{\text{eq}}}{\partial \epsilon }\right)\mathbf{v}_{\mathbf{p}}\mathbf{\cdot }\mathbf{\Lambda }^{\chi }\right).
\end{eqnarray}
Due to the fact that $\mathbf{\Omega}^{-\chi}\approx-\mathbf{\Omega}^{\chi}$, terms linear in $\mathbf{\Omega}^{\chi}$ vanish, and keeping up to order $\mathbf{\Omega}^2$, the longitudinal current is given by,
\begin{eqnarray}
\label{eq:jzmag}
J^z&{\cong}&J_{\text{eq}}^z+e\tau \int \frac{d^3p}{(2\pi )^3} \left(-\frac{\partial f_{\text{eq}}}{\partial \epsilon}\right) \times \nonumber \\
&&\left(\left(\left(v_{\mathbf{p}}^z\right)^2+\left(\frac{e
B}{c}\right)^2\left(v_{\mathbf{p}}^x\Omega _x^{\chi}+v_{\mathbf{p}}^y\Omega _y^{\chi}\right)^2\right)e E+\frac{\frac{e B}{c}\frac{e B}{c}\left(v_{\mathbf{p}}^x\Omega _x^{\chi}+v_{\mathbf{p}}^y\Omega _y^{\chi}+v_{\mathbf{p}}^z\Omega
_z^{\chi}\right)\left(v_{\mathbf{p}}^x\Omega _x^{\chi}+v_{\mathbf{p}}^y\Omega _y^{\chi}\right)}{1+\left(\tau  \frac{e B}{m c}\frac{2k_{\perp}^2}{m \epsilon
}\right)^2}e E\right).
\end{eqnarray}
\end{widetext}
To further simplify the integrand, we plug Eq.(\ref{eq: berry}) into Eq.(\ref{eq:jzmag}) and find
\begin{eqnarray}
\sigma _{z z}^{\text{dc}}(B)&\approx&2\sigma _{z z}^{\text{dc}}(B=0) \nonumber \\
&&+\frac{2\left(\hbar  \omega _c\right)^2}{\mu ^2}2\sigma _{z z}^{\text{dc}}(B=0),
\end{eqnarray}
where $\omega_c$ is the cyclotron frequency defined by $\omega_c=\frac{e B}{m c}$, which has no dependence on $v_z$. The factor of 2 comes from the pair of nodes with opposite chirality. 

\subsection{$\mathbf{B}=B \hat{z}, \mathbf{E}=0, \mathbf{\nabla} T=\nabla T \hat{z}$}
\label{subsec: B}

We now turn to the thermal transport with zero electrical field and a temperature gradient parallel to the magnetic field. In this case, Eq.(\ref{eq:Boltzmann2}) in the linear response regime becomes
\begin{widetext}
\begin{equation}
\label{eq:Boltzmann3}
f^{\chi }\approx f_{\text{eq}}-\left(1+\frac{e}{c}\mathbf{B}\cdot \mathbf{\Omega }^{\chi}\right)^{-1}\tau  \left(v_{\mathbf{p}}^z+\frac{e B}{c}\left(\mathbf{v}_{\mathbf{p}}\cdot
\mathbf{\Omega }^{\chi}\right)\right)\nabla _zT\frac{\partial f_{\text{eq}}}{\partial T}+\left(1+\frac{e}{c}\mathbf{B}\cdot \mathbf{\Omega }^{\chi}\right)^{-1}\tau \frac{e
B}{c} \left(v_{\mathbf{p}}^y\partial _{p_x}f^{\chi }-v_{\mathbf{p}}^x\partial _{p_y}f^{\chi }\right).
\end{equation}
To solve for $f^{\chi}$, we again make the ansatz
\begin{equation}
\label{eq:ansatz2}
f^{\chi }=f_{\text{eq}}-\left(1+\frac{e}{c}\mathbf{B}\cdot \mathbf{\Omega }^{\chi}\right)^{-1}\tau \left(v_{\mathbf{p}}^z+\frac{e}{c}\left(\mathbf{v}_{\mathbf{p}}\cdot
\mathbf{\Omega }^{\chi}\right)B\right)\nabla _zT\frac{\epsilon -\mu }{T}\left(-\frac{\partial f_{\text{eq}}}{\partial \epsilon }\right)+\left(-\frac{\partial
f_{\text{eq}}}{\partial \epsilon }\right)\mathbf{v}_{\mathbf{p}}\mathbf{\cdot }\mathbf{\Lambda }^{\chi }.
\end{equation}
By plugging Eq. (\ref{eq:ansatz2}) back into Eq. (\ref{eq:Boltzmann3}), we have
\begin{eqnarray}
\Lambda _x^{\chi }&=&\frac{\nabla _zT\frac{\epsilon -\mu }{T}\tau \frac{e B}{c}\tau  \frac{e B}{m c}\frac{2k_{\perp}^2}{m \epsilon }\Omega
_y^{\chi}+\nabla _zT\frac{\epsilon -\mu }{T}\left(1+\frac{e}{c}\mathbf{B}\cdot \mathbf{\Omega }^{\chi}\right)^{-1}\tau \frac{e B}{c}\left(\tau  \frac{e B}{m c}\frac{2k_{\perp}^2}{m
\epsilon }\right)^2\Omega _x^{\chi}}{\left(1+\frac{e}{c}\mathbf{B}\cdot \mathbf{\Omega }^{\chi}\right)^2+\left(\tau  \frac{e B}{m c}\frac{2k_{\perp}^2}{m
\epsilon }\right)^2}, \nonumber \\
 \Lambda _y^{\chi }&=&\frac{-\nabla _zT\frac{\epsilon -\mu }{T}\tau \frac{e B}{c}\tau  \frac{e B}{m c}\frac{2k_{\perp}^2}{m \epsilon }\Omega
_x+\nabla _zT\frac{\epsilon -\mu }{T}\left(1+\frac{e}{c}\mathbf{B}\cdot \mathbf{\Omega }^{\chi}\right)^{-1}\tau \frac{e B}{c}\left(\tau  \frac{e B}{m c}\frac{2k_{\perp}^2}{m
\epsilon }\right)^2\Omega _y^{\chi}}{\left(1+\frac{e}{c}\mathbf{B}\cdot \mathbf{\Omega }^{\chi}\right)^2+\left(\tau  \frac{e B}{m c}\frac{2k_{\perp}^2}{m
\epsilon }\right)^2}, \nonumber \\
 \Lambda _z^{\chi }&=&0.
\end{eqnarray}
The electrical and thermal currents are obtained as
\begin{eqnarray}
\mathbf{J}&=&-e\int \frac{d^3p}{(2\pi )^3}\left(\mathbf{v}_{\mathbf{p}}+\frac{e}{c}\left(\mathbf{v}_{\mathbf{p}}\cdot \mathbf{\Omega }^{\chi}\right)\mathbf{B}\right) \times \nonumber \\
&&\left(f_{\text{eq}}-\left(1+\frac{e}{c}\mathbf{B}\cdot
\mathbf{\Omega }^{\chi}\right)^{-1}\tau \left(v_{\mathbf{p}}^z+\frac{e}{c}\left(\mathbf{v}_{\mathbf{p}}\cdot \mathbf{\Omega }^{\chi}\right)B\right)\nabla _zT\frac{\epsilon -\mu
}{T}\left(-\frac{\partial f_{\text{eq}}}{\partial \epsilon }\right)+\left(-\frac{\partial f_{\text{eq}}}{\partial \epsilon }\right)\left(v_{\mathbf{p}}^x\Lambda
_x^{\chi }+v_{\mathbf{p}}^y\Lambda _y^{\chi }\right)\right) ,\\
\mathbf{J}^Q&=&-e\int \frac{d^3p}{(2\pi )^3}(\epsilon -\mu )\left(\mathbf{v}_{\mathbf{p}}+e \mathbf{E}\times \mathbf{\Omega }^{\chi}+\frac{e}{c}\left(\mathbf{v}_{\mathbf{p}}\cdot
\mathbf{\Omega }^{\chi}\right)\mathbf{B}\right)\times \nonumber \\
&&\left(f_{\text{eq}}-\left(1+\frac{e}{c}\mathbf{B}\cdot \mathbf{\Omega }^{\chi}\right)^{-1}\tau \left(v_{\mathbf{p}}^z+\frac{e}{c}\left(\mathbf{v}_{\mathbf{p}}\cdot
\mathbf{\Omega }^{\chi}\right)B\right)\nabla _zT\frac{\epsilon -\mu }{T}\left(-\frac{\partial f_{\text{eq}}}{\partial \epsilon }\right)+\left(-\frac{\partial
f_{\text{eq}}}{\partial \epsilon }\right)\left(v_{\mathbf{p}}^x\Lambda _x^{\chi }+v_{\mathbf{p}}^y\Lambda _y^{\chi }\right)\right).
\end{eqnarray}
Up to order $\mathbf{\Omega}^2$, the longitudinal current is given by
\begin{eqnarray}
\label{eq:jzmag2}
J^z&{\cong}&J_{\text{eq}}^z+e \tau \int \frac{d^3p}{(2\pi )^3}\left(-\frac{\partial f_{\text{eq}}}{\partial
\epsilon }\right) \times \nonumber \\
&&\left(\left(\left(v_{\mathbf{p}}^z\right)^2+\left(\frac{e
B}{c}\right)^2\left(v_{\mathbf{p}}^x\Omega _x^{\chi}+v_{\mathbf{p}}^y\Omega _y^{\chi}\right)^2\right)\nabla _zT\frac{\epsilon -\mu }{T}+\frac{\frac{e B}{c}\frac{e B}{c}\left(v_{\mathbf{p}}^x\Omega
_x^{\chi}+v_{\mathbf{p}}^y\Omega _y^{\chi}+v_{\mathbf{p}}^z\Omega _z^{\chi}\right)\left(v_{\mathbf{p}}^x\Omega _x^{\chi}+v_{\mathbf{p}}^y\Omega _y^{\chi}\right)}{1+\left(\tau  \frac{e
B}{m c}\frac{2k_{\perp}^2}{m \epsilon }\right)^2}\nabla _zT\frac{\epsilon -\mu }{T}\right).
\end{eqnarray}
\end{widetext}
Notice that we use the fact that
$\Omega _x^{\chi}\propto k_x, \Omega _y^{\chi}\propto k_y, \Omega _z^{\chi}\propto k_z,
v_{\mathbf{p}}^x\propto k_x, v_{\mathbf{p}}^y\propto k_y,v_{\mathbf{p}}^z\propto k_z$
to simplify the integrand. To further simplify Eq.(\ref{eq:jzmag2}), we use Eq.(\ref{eq: berry}) and find
\begin{equation}
L_{12}^{z z}(B)=2L_{12}^{z z}(B=0)+\frac{2\left(\hbar  \omega _c\right)^2}{\mu ^2}2L_{12}^{z
z}(B=0).
\end{equation}
Likewise, the longitudinal thermal current is derived as
\begin{widetext}
\begin{eqnarray}
J_z^Q&{\cong}&J_{\text{eq}, z}^Q+e \tau \int \frac{d^3p}{(2\pi )^3}(\epsilon -\mu )\left(-\frac{\partial f_{\text{eq}}}{\partial \epsilon }\right) \times \nonumber \\
&&\left(\left(\left(v_{\mathbf{p}}^z\right)^2+\left(\frac{e B}{c}\right)^2\left(v_{\mathbf{p}}^x\Omega _x^{\chi}+v_{\mathbf{p}}^y\Omega _y^{\chi}\right)^2\right)\nabla _zT\frac{\epsilon
-\mu }{T}+\frac{\frac{e B}{c}\frac{e B}{c}\left(v_{\mathbf{p}}^x\Omega _x^{\chi}+v_{\mathbf{p}}^y\Omega _y^{\chi}+v_{\mathbf{p}}^z\Omega _z^{\chi}\right)\left(v_{\mathbf{p}}^x\Omega _x^{\chi}+v_{\mathbf{p}}^y\Omega
_y^{\chi}\right)}{1+\left(\tau  \frac{e B}{m c}\frac{2k_{\perp}^2}{m \epsilon }\right)^2}\nabla _zT\frac{\epsilon -\mu }{T}\right), \nonumber \\
\end{eqnarray}
\end{widetext}
and
\begin{equation}
L_{22}^{z z}(B)=2L_{22}^{z z}(B=0)+\frac{2\left(\hbar  \omega _c\right)^2}{\mu ^2}2L_{22}^{zz}(B=0).
\end{equation}
We find that the longitudinal component of $L_{12}$ and $L_{22}$ have the same dependence on cyclotron frequency and chemical potential as electrical conductivity, which means the Mott relation and Wiedemann-Franz law are preserved. The thermal conductivity has the same dependence on magnetic field strength and zero field counterpart as $L_{11}$, $L_{12}$ and $L_{22}$:
\begin{equation}
\kappa _{z z}(B) \approx 2\kappa_{z z}(B=0)+\frac{2\left(\hbar  \omega _c\right)^2}{\mu ^2}2\kappa_{z z}(B=0).
\end{equation}

\subsection{$\mathbf{B}=B \hat{z}, \mathbf{E}=E \hat{x}, \mathbf{\nabla} T=0$}
\label{sec:C}

For probing fields ($\mathbf{E}$ and $\mathbf{\nabla} T$) being perpendicular to magnetic field, we first consider the case with zero temperature gradient. Following previous sections, the solution for distribution function reads
\begin{equation}
\label{eq:ansatz3}
f^{\chi }=f_{\text{eq}}+\left(1+\frac{e}{c}\mathbf{B}\cdot \mathbf{\Omega }^{\chi}\right)^{-1}\tau  e E\nabla _{p_x}f_{\text{eq}}+\left(-\frac{\partial f_{\text{eq}}}{\partial
\epsilon }\right)\mathbf{v}_{\mathbf{p}}\mathbf{\cdot }\mathbf{\Lambda }^{\chi },
\end{equation}
with
\begin{eqnarray}
 \Lambda _x^{\chi }&=&\frac{\left(1+\frac{e}{c}\mathbf{B}\cdot \mathbf{\Omega }^{\chi}\right)^{-1}\tau  \left(\tau \frac{e B}{m c}\frac{2k_{\perp}^2}{m
\epsilon }\right)^2e E}{\left(1+\frac{e}{c}\mathbf{B}\cdot \mathbf{\Omega }^{\chi}\right)^2+\left(\tau  \frac{e B}{m c}\frac{2k_{\perp}^2}{m \epsilon
}\right)^2}, \nonumber \\
 \Lambda _y^{\chi }&=&\frac{-\tau ^2\frac{e B}{m c}\frac{2k_{\perp}^2}{m \epsilon }e E}{\left(1+\frac{e}{c}\mathbf{B}\cdot \mathbf{\Omega }^{\chi}\right)^2+\left(\tau
\frac{e B}{m c}\frac{2k_{\perp}^2}{m \epsilon }\right)^2}, \nonumber \\
 \Lambda _z^{\chi }&=&0.
\end{eqnarray}
The electrical current and thermal current are calculated in the linear response regime as
\begin{widetext}
\begin{eqnarray}
\mathbf{J}&=&-e\int \frac{d^3p}{(2\pi )^3}\left(1+\frac{e}{c}\mathbf{B}\cdot \mathbf{\Omega }^{\chi}\right)\dot{\mathbf{r}}f^{\chi } \nonumber \\
&\approx& -e\int \frac{d^3p}{(2\pi )^3}\left(\mathbf{v}_{\mathbf{p}}+\frac{e}{c}\left(\mathbf{v}_{\mathbf{p}}\cdot \mathbf{\Omega }^{\chi}\right)\mathbf{B}\right)\times \nonumber \\ 
&&\left(f_{\text{eq}}-v^x_{\mathbf{p}}\frac{\left(1+\frac{e}{c}\mathbf{B}\cdot
\mathbf{\Omega }^{\chi}\right)^{-1}\tau  \left(1+\frac{e}{c}\mathbf{B}\cdot \mathbf{\Omega }^{\chi}\right)^2}{\left(1+\frac{e}{c}\mathbf{B}\cdot \mathbf{\Omega }^{\chi}\right)^2+\left(\tau
 \frac{e B}{m c}\frac{2k_{\perp}^2}{m \epsilon }\right)^2} \left(-\frac{\partial f_{\text{eq}}}{\partial \epsilon }\right)e E-v^y_{\mathbf{p}}\frac{\tau
^2\frac{e B}{m c}\frac{2k_{\perp}^2}{m \epsilon }}{\left(1+\frac{e}{c}\mathbf{B}\cdot \mathbf{\Omega }^{\chi}\right)^2+\left(\tau \frac{e
B}{m c}\frac{2k_{\perp}^2}{m \epsilon }\right)^2}\left(-\frac{\partial f_{\text{eq}}}{\partial \epsilon }\right)e E\right) \nonumber \\
&&-e^2\int
\frac{d^3p}{(2\pi )^3} \mathbf{E}\times \mathbf{\Omega }\mathbf{ }f_{\text{eq}}, \\
\mathbf{J}^Q&=&\int \frac{d^3p}{(2\pi )^3}(\epsilon -\mu )\left(1+\frac{e}{c}\mathbf{B}\cdot \mathbf{\Omega }^{\chi}\right)\dot{\mathbf{r}}f^{\chi } \nonumber \\
&\approx& \int \frac{d^3p}{(2\pi )^3}(\epsilon -\mu )\left(\mathbf{v}_{\mathbf{p}}+\frac{e}{c}\left(\mathbf{v}_{\mathbf{p}}\cdot \mathbf{\Omega }^{\chi}\right)\mathbf{B}\right) \times \nonumber \\
&&\left(f_{\text{eq}}-v^x_{\mathbf{p}}\frac{\left(1+\frac{e}{c}\mathbf{B}\cdot
\mathbf{\Omega }^{\chi}\right)^{-1}\tau  \left(1+\frac{e}{c}\mathbf{B}\cdot \mathbf{\Omega }^{\chi}\right)^2}{\left(1+\frac{e}{c}\mathbf{B}\cdot \mathbf{\Omega }^{\chi}\right)^2+\left(\tau
 \frac{e B}{m c}\frac{2k_{\perp}^2}{m \epsilon }\right)^2} \left(-\frac{\partial f_{\text{eq}}}{\partial \epsilon }\right)e E-v^y_{\mathbf{p}}\frac{\tau
^2\frac{e B}{m c}\frac{2k_{\perp}^2}{m \epsilon }}{\left(1+\frac{e}{c}\mathbf{B}\cdot \mathbf{\Omega }^{\chi}\right)^2+\left(\tau \frac{e
B}{m c}\frac{2k_{\perp}^2}{m \epsilon }\right)^2}\left(-\frac{\partial f_{\text{eq}}}{\partial \epsilon }\right)e E\right) \nonumber \\
&&+e\int
\frac{d^3p}{(2\pi )^3}(\epsilon -\mu ) \mathbf{E}\times \mathbf{\Omega }^{\chi}\mathbf{ }f_{\text{eq}}.
\end{eqnarray}
\end{widetext}
First we consider the longitudinal components. By expansion up to linear order in $\frac{e}{c}\mathbf{B}\cdot \mathbf{\Omega }$ and resorting to $\mathbf{\Omega }^{\chi }\approx -\mathbf{\Omega }^{-\chi
}$, we could skip linear terms in $\mathbf{\Omega}^{\chi}$ due to summation over opposite chirality contribution. The longitudinal conductivity is evaluated as
\begin{eqnarray}
\sigma _{x x}^{\text{dc}}(B)&=&\frac{3\left(2\omega _c\tau -\frac{\text{arcsinh}\left(2\omega _c\tau \right)}{\sqrt{1+\left(2\omega _c\tau \right)^2}}\right)}{2\left(2\omega
_c\tau \right)^3}2\sigma _{x x}^{\text{dc}}(B=0) \nonumber \\
&\approx& \left(1-\frac{16 \left(\omega _c\tau \right)^2}{5}\right)2\sigma _{x x}^{\text{dc}}(B=0),
\end{eqnarray}
which turns out to decrease with the growth of magnetic field strength similar to the single Weyl point\cite{Rexprb2014}. Likewise, the transverse conductivity can be derived as
\begin{eqnarray}
\sigma _{y x}^{\text{dc}}&=&\frac{3\pi }{8}\frac{\left(-2+\left(2\tau  \omega _c\right)^2+\frac{2}{\sqrt{1+\left(2\tau  \omega _c\right)^2}}\right)}{\left(2\tau
 \omega _c\right)^3}2\sigma _{x x}^{\text{dc}}(B=0) \nonumber \\
 &&+\sigma _{y x}^A \nonumber \\
&\approx& \frac{9\pi  \left(\omega _c\tau \right)}{16}2\sigma _{x x}^{\text{dc}}(B=0)-2\frac{e^2}{h}\frac{2 g B}{\pi}.
\end{eqnarray}

\subsection{$\mathbf{B}=B \hat{z}, \mathbf{E}=0, \mathbf{\nabla} T=\nabla T \hat{x}$}
\label{subsec: D}

Finally, we investigate the case with zero electrical field and a perpendicular temperature gradient to the magnetic field. The steady state distribution function is shown to be
\begin{widetext}
\begin{eqnarray}
f^{\chi }&=&f_{\text{eq}}-\left(1+\frac{e}{c}\mathbf{B}\cdot \mathbf{\Omega }^{\chi}\right)^{-1}\tau  v_x\nabla _xf_{\text{eq}}+\left(-\frac{\partial f_{\text{eq}}}{\partial
\epsilon }\right)\mathbf{v}\mathbf{\cdot }\mathbf{\Lambda }^{\chi }, \nonumber \\
\end{eqnarray}
with
\begin{eqnarray}
 \Lambda_x^{\chi }&=&\left(1+\frac{e}{c}\mathbf{B}\cdot \mathbf{\Omega }\right)^{-1}\tau  \frac{\left( \tau \frac{e B}{m c} \frac{2k_{\perp}^2}{m
\epsilon }\right)^2\nabla _xT \frac{\epsilon -\mu }{T}}{\left(1+\frac{e}{c}\mathbf{B}\cdot \mathbf{\Omega }\right)^2+\left(\tau \frac{e B}{m c} \frac{2k_{\perp}^2}{m
\epsilon }\right)^2}, \nonumber \\
 \Lambda_y^{\chi }&=&-\frac{ \tau ^2\frac{e B}{m c} \frac{2k_{\perp}^2}{m \epsilon }\nabla _xT \frac{\epsilon -\mu }{T}}{\left(1+\frac{e}{c}\mathbf{B}\cdot
\mathbf{\Omega }\right)^2+\left(\tau \frac{e B}{m c} \frac{2k_{\perp}^2}{m \epsilon }\right)^2}, \nonumber \\
\Lambda_z^{\chi }&=&0.
\end{eqnarray}
The electrical and thermal currents are found to be
\begin{eqnarray}
\mathbf{J}&=&-e\int \frac{d^3p}{(2\pi )^3}\left(1+\frac{e}{c}\mathbf{B}\cdot \mathbf{\Omega }^{\chi}\right)\dot{\mathbf{r}}f^{\chi } \nonumber \\
&=&-e\int \frac{d^3p}{(2\pi )^3}\left(\mathbf{v}_{\mathbf{p}}+\frac{e}{c}\left(\mathbf{v}_{\mathbf{p}}\cdot \mathbf{\Omega }^{\chi}\right)\mathbf{B}\right)f_{\text{eq}} \nonumber \\
&&+e\int \frac{d^3p}{(2\pi )^3}\left(\mathbf{v}_{\mathbf{p}}+\frac{e}{c}\left(\mathbf{v}_{\mathbf{p}}\cdot \mathbf{\Omega }^{\chi}\right)\mathbf{B}\right) \left(-\frac{\partial f_{\text{eq}}}{\partial \epsilon}\right) \times \nonumber \\
&&\left(v^x_{\mathbf{p}}\left(1+\frac{e}{c}\mathbf{B}\cdot \mathbf{\Omega }^{\chi}\right)^{-1}\tau \frac{ \nabla _xT \frac{\epsilon -\mu }{T}\left(1+\frac{e}{c}\mathbf{B}\cdot \mathbf{\Omega }^{\chi}\right)^2}{\left(1+\frac{e}{c}\mathbf{B}\cdot \mathbf{\Omega }^{\chi}\right)^2+\left(\tau \frac{e B}{m c} \frac{2k_{\perp}^2}{m \epsilon }\right)^2}+v^y_{\mathbf{p}}\frac{ \tau ^2\frac{e
B}{m c} \frac{2k_{\perp}^2}{m \epsilon }\nabla _xT \frac{\epsilon -\mu }{T}}{\left(1+\frac{e}{c}\mathbf{B}\cdot \mathbf{\Omega }^{\chi}\right)^2+\left(\tau \frac{e B}{m c} \frac{2k_{\perp}^2}{m \epsilon }\right)^2}\right), \\
\mathbf{J}^Q&=&-e\int \frac{d^3p}{(2\pi )^3}(\epsilon -\mu )\left(1+\frac{e}{c}\mathbf{B}\cdot \mathbf{\Omega }^{\chi}\right)\dot{\mathbf{r}}f^{\chi } \nonumber \\
&=&-e\int \frac{d^3p}{(2\pi )^3}(\epsilon -\mu )\left(\mathbf{v}_{\mathbf{p}}+\frac{e}{c}\left(\mathbf{v}_{\mathbf{p}}\cdot \mathbf{\Omega }^{\chi}\right)\mathbf{B}\right)f_{\text{eq}} \nonumber \\
&&+e\int \frac{d^3p}{(2\pi )^3}(\epsilon -\mu )\left(\mathbf{v}_{\mathbf{p}}+\frac{e}{c}\left(\mathbf{v}_{\mathbf{p}}\cdot \mathbf{\Omega }^{\chi}\right)\mathbf{B}\right) \left(-\frac{\partial f_{\text{eq}}}{\partial \epsilon
}\right) \times \nonumber \\
&&\left(v^x_{\mathbf{p}}\left(1+\frac{e}{c}\mathbf{B}\cdot
\mathbf{\Omega }^{\chi}\right)^{-1}\tau \frac{\nabla _xT \frac{\epsilon -\mu }{T}\left(1+\frac{e}{c}\mathbf{B}\cdot \mathbf{\Omega }^{\chi}\right)^2}{\left(1+\frac{e}{c}\mathbf{B}\cdot
\mathbf{\Omega }^{\chi}\right)^2+\left(\tau \frac{e B}{m c} \frac{2k_{\perp}^2}{m \epsilon }\right)^2}+v^y_{\mathbf{p}}\frac{\tau ^2\frac{e
B}{m c} \frac{2k_{\perp}^2}{m \epsilon }\nabla _xT \frac{\epsilon -\mu }{T}}{\left(1+\frac{e}{c}\mathbf{B}\cdot \mathbf{\Omega }^{\chi}\right)^2+\left(\tau
\frac{e B}{m c} \frac{2k_{\perp}^2}{m \epsilon }\right)^2}\right).
\end{eqnarray}
Again we expand up to linear order in $\frac{e}{c}\mathbf{B}\cdot \mathbf{\Omega}^{\chi}$ and resorting to $\mathbf{\Omega }^{\chi }\approx -\mathbf{\Omega }^{-\chi
}$, the longitudinal and transverse components of thermoelectric response coefficients in the weak field regime are obtained as
\begin{eqnarray}
L_{x x}^{12}(B)&\approx& \left(1-\frac{16 \left(\omega _c\tau \right)^2}{5}\right)2L_{x x}^{12}(B=0), \\
L_{x x}^{22}(B)&\approx& \left(1-\frac{16 \left(\omega _c\tau \right)^2}{5}\right)2L_{x x}^{22}(B=0), \\
L_{y x}^{12}(B)&\approx& \frac{9\pi  \omega _c\tau}{16}2L_{x x}^{12}(B=0), \\
L_{y x}^{22}(B)&\approx& \frac{9\pi  \omega _c\tau }{16} 2L_{x x}^{22}(B=0)-\frac{2\pi }{3h}k_B^2T(2 g B), \nonumber \\
\end{eqnarray}
\end{widetext}
which also have the same dependence on $\omega_c$ and $\mu$ as the electrical conductivity obtained in \ref{sec:C}. The thermal conductivity are found to be
\begin{eqnarray}
\kappa _{x x}(B) &\approx& 2\kappa_{x x}(B=0)-\frac{16\left(\omega _c \tau\right)^2}{5}2\kappa_{x x}(B=0), \nonumber \\
\kappa _{x y}(B) &\approx& \frac{9 \pi \omega_c \tau}{16} 2\kappa_{x x}(B=0)-\frac{2\pi }{3h}k_B^2T(2 g B). \nonumber \\
\end{eqnarray}
\begin{table*}[h!]
\begin{center}
\begin{tabular*}{\textwidth}{| P{0.3\textwidth} | P{0.045\textwidth} | P{0.31\textwidth} | P{0.31\textwidth} |}
\hline
&$\alpha \beta$& $\sigma_{\alpha \beta}$ & $\kappa_{\alpha \beta}$ \\
\hline
$\mathbf{B}=B \hat{z}, \mathbf{E}=E \hat{z}\ \mathrm{or}\ \mathbf{\nabla} T=\nabla T \hat{z}$ &$z z$&$(1+\frac{2\left(\hbar  \omega _c\right)^2}{\mu ^2})2\sigma _{z z}^{\text{dc}}(B=0)$ & $(1+\frac{2\left(\hbar  \omega _c\right)^2}{\mu ^2})2\kappa_{z z}(B=0)$\\
\cline{2-4}
&$x y$&$0$ & $0$\\
\hline
$\mathbf{B}=B \hat{z}, \mathbf{E}=E \hat{x}\ \mathrm{or}\ \mathbf{\nabla} T=\nabla T \hat{x}$ &$x x$&$\left(1-\frac{16 \left(\omega _c\tau \right)^2}{5}\right)2\sigma _{x x}^{\text{dc}}(B=0)$&$\left(1-\frac{16 \left(\omega _c\tau \right)^2}{5}\right)2\kappa_{x x}(B=0)$\\
\cline{2-4}
&$x y$&$\frac{9\pi  \left(\omega _c\tau \right)}{16}2\sigma _{x x}^{\text{dc}}(B=0)-2\frac{e^2}{h}\frac{2 g B}{\pi}$&$\frac{9 \pi \omega_c \tau}{16} 2\kappa_{x x}(B=0)-\frac{2\pi }{3h}k_B^2T(2 g B)$\\
\hline
\end{tabular*}
\end{center}
\caption{Thermoelectric transport coefficients in the semi-classical regime $\mu \gg \hbar \omega_c$ with static magnetic field applied in $z$ direction. Here $\sigma_{\alpha \beta}$ is the electrical conductivity and $\kappa_{\alpha \beta}$ the thermal conductivity with spatial components labeled by $\alpha,\beta$.}
\label{table:withfield}
\end{table*}
\newpage
\subsection{Anomalous chiral current and chiral magnetic effects}

In the above sections, we neglected the internode scattering effects, which turns out to be important for chiral anomaly related transport\cite{DTSonprb2013,Spivak2015,Yip2015}.  In the regime where the elastic intranode scattering time $\tau$ is the shortest relaxation time in the problem, one can neglect the intranode anisotropy of the electron distribution and describe the system with an energy dependent distribution function $f_{\epsilon }^{\chi }(\mathbf{r},t)$\cite{Spivak2015}. The chiral anomaly effect is summarized in the kinetic equation
\begin{widetext}
\begin{equation}
\label{eq:kinetic}
\frac{\partial }{\partial t}f_{\epsilon }^{\chi }(\mathbf{r},t)+\nabla _{\mathbf{r}}\cdot \mathbf{j}^{\chi }(\epsilon ,\mathbf{r},t)-\frac{k^{\chi }}{g^{\chi
}(\epsilon )}\frac{e}{(2\pi  \hbar )^2c}\mathbf{B}\mathbf{\cdot }\frac{\partial f_{\epsilon }^{\chi }(\mathbf{r},t)}{\partial \mathbf{r}}=-\frac{\delta  f_{\epsilon
}^{\chi }(\mathbf{r},t)}{\tau _I},
\end{equation}
\end{widetext}
derived in Ref. [\onlinecite{DTSonprb2013, Spivak2015}], where 
\begin{equation}
k^{\chi}=\frac{1}{2 \pi  \hbar}\int \mathbf{\Omega }^{\chi } \cdot d \mathbf{S}
\end{equation}
is the quantized flux (or chiral charge) of the Berry curvature through the constant energy surface. $k^{\chi}=2 \chi$ for the double-Weyl node with given chirality $\chi$,   \(\tau _I\) is the internode relaxation time, and \(g^{\chi }(\epsilon )\) is the modified density of states with chirality $\chi $ given by,
\begin{equation}
g^{\chi }(\epsilon )=\int \frac{d^3p}{(2\pi  \hbar )^3}\left(1+\frac{e}{c}\mathbf{B}\cdot \mathbf{\Omega }^{\chi }\right)\delta \left(\epsilon _{\mathbf{p}}-\epsilon
\right).
\end{equation}
For simplicity, we assumed \(\mu >0\) and only consider the conduction band and,
\begin{eqnarray}
\label{eq:chiralcurrent}
\mathbf{j}^{\chi }(\epsilon ,\mathbf{r},t)&=&f_{\epsilon }^{\chi }(\mathbf{r},t)\int \frac{d^3p}{(2\pi  \hbar )^3}\nonumber \\
&&\times \left(\mathbf{v}_{\mathbf{p}}+e \mathbf{E}\times \mathbf{\Omega
}^{\chi }+\frac{e}{c}\left(\mathbf{v}_{\mathbf{p}}\cdot \mathbf{\Omega }^{\chi }\right)\mathbf{B}\right) \delta \left(\epsilon _{\mathbf{p}}-\epsilon \right),\nonumber \\
\end{eqnarray}
is the particle flux density at a given energy $\epsilon $ and chirality $\chi $. In the spatially homogeneous case, the deviation of the steady state distribution function from equilibrium is solved in the linear response regime of applied field ($\mathbf{E}$ and $\mathbf{\nabla}T$, respectively) as,
\begin{eqnarray}
\delta  f_{\epsilon }^{\chi }&=&-\frac{k^{\chi }}{g^{\chi }(\epsilon )}\frac{e^2\tau }{(2\pi  \hbar )^2c}(\mathbf{E\cdot B})\frac{\partial f_{\epsilon
}^{\chi }}{\partial \epsilon }, \\
\delta  f_{\epsilon }^{\chi }&=&-\frac{k^{\chi }}{g^{\chi }(\epsilon )}\frac{e \tau }{(2\pi  \hbar )^2c}\mathbf{B}\mathbf{\cdot }\frac{(\epsilon -\mu )(-\nabla
T)}{T}\frac{\partial f_{\epsilon }^{\chi }}{\partial \epsilon },
\end{eqnarray}
with \(g^{\chi }(\epsilon )\approx \frac{m |\epsilon |}{4\pi  \hbar ^3v_z}\). For \(\frac{\mu }{k_BT}\gg 1\), the thermoelectric coefficients are summarized in Table. \ref{table:anomaly}. In the non degenerate limit \(\frac{\mu }{k_BT}\ll 1\), inelastic scattering of electrons may no longer be ignored and the approach based on quasi-classical Boltzmann equation could break down at low energies\cite{Spivak2015}.

\begin{figure}[!t]
\epsfig{file=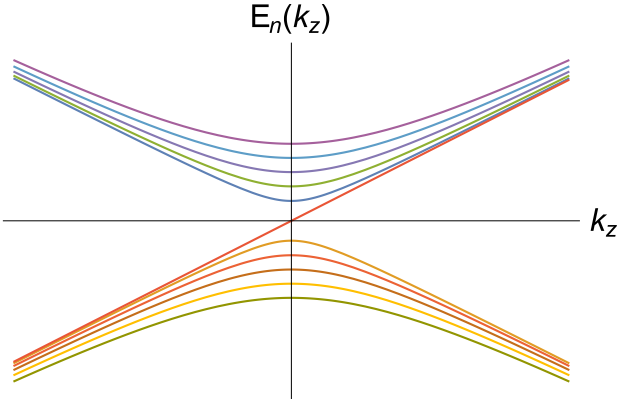,clip=0.1,width=0.95\linewidth,angle=0}
\caption{Landau level spectrum for double-Weyl fermion under magnetic field. The lowest Landau levels are both chiral and doubly degenerate.}
\label{fig:landaulevel}
\end{figure}

The above results are valid for $\mu \gg \hbar \omega_c$ with TRS broken. In the opposite ultraquantum limit \(\omega _c\tau _{\text{tr}}\gg1,\mu
<\hbar \omega _c\), we need to consider the lowest Landau level contribution to the magnetoconductivity in TRS preserving case. The Landau level spectrum is given by,
\begin{eqnarray}
 E_n&=&\pm \sqrt{n(n-1)(\hbar \omega _c)^2+v_z^2\left(\hbar k_z\right)^2}, n\geq 2. \nonumber \\
 E_{0,1}&=&\chi v_z \hbar k_z,
\end{eqnarray}
which turns out to be analogous to 2D bilayer graphene\cite{McCann2013}.  In particular, the lowest Landau level has doubly degenerate chiral modes (Fig. \ref{fig:landaulevel}). According to the analysis of Ref. [\onlinecite{DTSonprb2013, Nielsen1983}], the anomalous
magnetoconductivity for \(\frac{\mu }{k_BT}\gg 1\) is a direct consequence of the anomalous \(\mathbf{E\cdot B}\) term in linear response, which is calculated as
\begin{equation}
\label{eq:sigmaLLL}
\sigma _{z z}=2\frac{e^3 B v_z}{4\pi^2 \hbar^2 c}\tau _I.
\end{equation}
The factor of 2 compared to the Weyl point with linear dispersion results from the double degeneracy (chiral charge) of the chiral modes. Eq.(\ref{eq:sigmaLLL}) could also be obtained from the kinetic equation Eq.(\ref{eq:kinetic}) with the density of states for lowest Landau levels given by
\begin{equation}
g^{\chi }(\epsilon )=\frac{1}{L_x L_y L_z}\sum _{k_z} \delta \left(\epsilon -v_zk_z\right)2N_{\phi }=\frac{e B}{2\pi ^2\hbar ^2c v_z},
\end{equation}
where $N_\phi=\frac{e B L_x L_y}{2 \pi \hbar c}$ is the number of single particle states in a given Landau level. Similarly, other thermoelectric coefficients are obtained in Table. \ref{table:anomaly}.

Another related phenomenon due to the role of double-Weyl points as source and sink for Berry curvature is the chiral magnetic effect\cite{DTSonprl2012, Vilenkinprd1980, Fukushimaprd2008}. If one looks at the right hand side of Eq. (\ref{eq:chiralcurrent}), the first and second terms give rise to normal and anomalous current respectively, while the third term causes the chiral magnetic effect. The chiral current for each node is given by
\begin{equation}
\mathbf{J}_C^{\chi }=e \int \frac{d^3\mathbf{p}}{(2\pi \hbar)^3}\frac{e}{c}\left(\mathbf{\Omega }_{\mathbf{p}}^{\chi }\cdot \mathbf{v}_{\mathbf{p}}\right)\mathbf{B}f_{\mathbf{p}}^{\chi },
\end{equation}
at zero temperature,
\begin{equation}
J_C^{\chi , z}=\frac{2\chi  e^2}{4\pi ^2 \hbar^2 c}\mu _{\chi }B_z.
\end{equation}
Notice that the magnetic field has to point in the z-direction to preserve (i.e., not split) the double weyl nodes. This means the chemical potential imbalance between double-Weyl nodes will drive current flow even without electrical field. The proportional constant doubles the contribution from linearly dispersing Weyl nodes.

\begin{table*}[h!]
\begin{center}
\begin{tabular*}{\textwidth}{| P{0.2146\textwidth} | P{0.2\textwidth} | P{0.35\textwidth} | P{0.2\textwidth}|}
\hline
&$\sigma_{z z}$ & $\kappa_{z z}$ & $S_z$ \\
\hline
$\mu \gg \hbar \omega_c, \frac{\mu}{k_B T} \gg 1$&$2\frac{e^2B^2}{\pi ^2\hbar  c}\frac{e^2}{m \mu }\frac{v_z}{c}\tau _I$ & $2\frac{e^2B^2}{\pi ^2\hbar  c}\frac{v_z}{m c}\tau _I \frac{\pi ^2}{3}k_B \left(\frac{k_BT}{\mu }-\frac{\pi ^2}{3}\left(\frac{k_BT}{\mu}\right)^3\right)$ & $-\frac{\pi ^2k_B^2 T}{3 e \mu }$\\
\hline
$\mu \ll \hbar \omega_c, \frac{\mu}{k_B T} \gg 1$&$2\frac{e^3 B v_z}{4\pi^2 \hbar^2 c}\tau _I$ & $\frac{e v_zB}{2\pi ^2\hbar ^2c}\tau _I\frac{\pi ^2}{3}k_B^2T$ & $0$\\
\hline
\end{tabular*}
\end{center}
\caption{Anomalous thermoelectric transport coefficients with static magnetic field applied in $z$ direction. Here $\sigma$ is the electrical conductivity, $\kappa$ the thermal conductivity, and $S$ the Seebeck coefficient.}
\label{table:anomaly}
\end{table*}


\section{CONCLUSION AND DISCUSSION}
\label{sect:summary}

In our work, we have investigated the electronic contribution to the thermoelectric properties of the double Weyl semimetal system mainly via the semiclassical Boltzmann transport formalism. We found that the in-plane and out-of-plane longitudinal transport coefficients have different dependence on the temperature due to the anisotropy in the band dispersion. The anomalous transport coefficients are doubled due to the doubling of topological charge. We also considered both the electron-electron interaction effects and the diffusive transport with short-ranged disorder potential. By doping away from the nodal point, the diffusive thermoelectric transport quantities have an interesting directional dependence on both the chemical potential and model parameters. We confirmed our results at zero temperature by comparison to the diagrammatic approach including the vertex correction by a ladder sum. When approaching the charge neutral point (low doping), the an-isotropically screened electron-electron interaction leads to linear dependence of relaxation rate on temperature (up to logarithmic corrections). As a result, the temperature dependence of thermoelectric properties reduces by one power of $k_B T$ compared to the noninteracting case. Our dissipative thermoelectric transport results under different relaxation processes indicate that the double Weyl node can be distinguished from the linearly dispersing Weyl node by examining the difference in temperature and chemical potential dependence between the in-plane and out-of-plane response coefficients.

By applying a static magnetic field in the linear dispersing direction, we find positive electrical (thermal) magneto-conductivity if the electrical field (temperature gradient) is in parallel with the magnetic field similar to the case of single Weyl point. When the electrical field (temperature gradient) is applied perpendicular to the magnetic field, we find that the in-plane longitudinal electrical (thermal) conductivity pick up a negative contribution quadratic in magnetic field strength, while the transverse components receives odd power corrections in magnetic field strength. In our work, we did not emphasize the effect of spatial anisotropy in scattering events, especially when the probing field ($\mathbf{E}$ and $\mathbf{\nabla} T$) are perpendicular to each other. In this case, the collision term will become more complicated and the distribution function can not be readily solved in general. 

In considering the magnetic field effects, the semiclassical approach we used is valid when the chemical potential is away from the band touching point. In the ultraquantum limit $(\mu \ll \hbar \omega_c)$, the semi-classical approach breaks down. Due to the chiral nature of doubly degenerate lowest Landau levels, the chiral anomaly contribution to the longitudinal thermoelectric coefficients doubles that of the single Weyl point. 

In real materials, $\mathrm{HgCr_2Se_4}$ has two kinds of band crossings close to the Fermi level ($\Delta E\sim0.01\mathrm{eV}$), i.e. a closed loop surrounding the $\Gamma$ point and two double-Weyl points located along the $\Gamma-Z$ line\cite{Xuprl2011}. Both of them will affect the measurement of thermoelectric transport properties. But it turns out that the loop crossing is not as stable as the nodal point and can be eliminated by changing the crystal symmetry\cite{Xuprl2011}. On the other hand, the energy separation of double-Weyl points in $\mathrm{SrSi_2}$ ($\Delta E\sim0.1\mathrm{eV}$)\cite{ShinMing2015} suggest itself to be a promising platform for chiral anomaly transport measurement.

To sum up, our results might be applied in the search for real materials that possess double-Weyl points near the Fermi surface, as well as a motivation for the experimental study of the thermoelectric properties of this particular band dispersion.  Our main results are summarized in Tables I, II and III.

\section{Acknowledgment}  We gratefully acknowledge financial support from ARO grant W911NF-14-1-0579 and NSF DMR-1507621. We thank Ki-Seok Kim, Hsin-Hua Lai, Shao-Kai Jian, Pavan Hosur, Yang Gao, Qian Niu, Pontus Laurell, Rex Lundgren, and Chungwei Lin for helpful discussions. 


%


\appendix

\section{DIAGRAMMATIC APPROACH OF DIFFUSIVE TRANSPORT}
\label{sec: diffusive diagram}

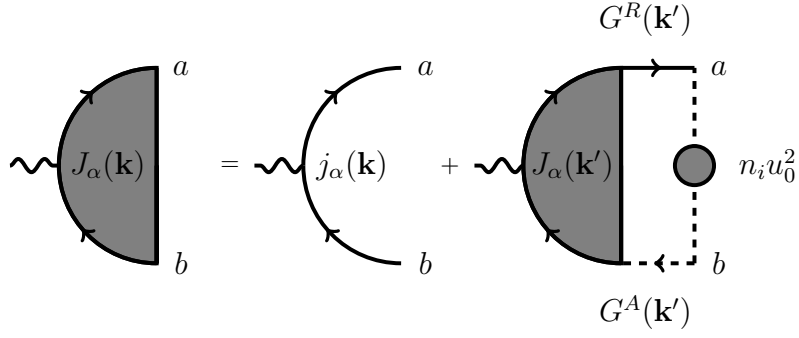
\begin{figure*}[!htb]
\centering
\vspace{2em}

\begin{tikzpicture}[line width=1.5 pt, scale=1.3]
        \filldraw[fill=gray] (1,-1)--(1,1) arc (90:270:1)--(1,0);
	\draw[fermion] (1,-1) arc (-90:-180:1);
	\draw[fermion] (0,0) arc(180:90:1);
	\draw[vector] (180:0.5)--(0,0);
	\node at (0:0.5) {\large $J_{\alpha}(\mathbf{k})$};
	\draw[fermionnoarrow] (1,1) -- (1,-1);
	\node at (1.25,1) {\large $a$};
	\node at (1.25,-1) {\large $b$};
	\node at (1.75,0) {$=$};
	
	\draw[fermion] (3.5,-1) arc (-90:-180:1);
	\draw[fermion] (2.5,0) arc(180:90:1);
	\draw[vector] (2.5,0)--(2,0);
	\node at (3, 0) {\large $j_{\alpha}(\mathbf{k})$};
	\node at (3.75,1) {\large $a$};
	\node at (3.75,-1) {\large $b$};
	\node at (4,0) {$+$};
	
	\filldraw[fill=gray] (5.75,-1)--(5.75,1) arc (90:270:1)--(5.75,0);
	\draw[fermion] (5.75,-1) arc (-90:-180:1);
	\draw[fermion] (4.75,0) arc(180:90:1);
	\draw[vector] (4.75,0)--(4.25,0);
	\node at (5.25, 0) {\large $J_{\alpha}(\mathbf{k}')$};
	\draw[fermionnoarrow] (5.75,1) -- (5.75,-1);
	\draw[fermion] (5.75,1) -- (6.5,1);
	\node at (6, 1.5) {\large $G^R(\mathbf{k}')$};
	\draw[scalar] (6.5,-1) -- (5.75,-1);
	\node at (6, -1.5) {\large $G^A(\mathbf{k}')$};
	\draw[scalarnoarrow] (6.5,-1) -- (6.5,1);
	\filldraw[fill=gray](6.5,0) circle(0.2);
	\node at (7.25,0) {\large $n_i u_0^2$};
	\node at (6.75,1) {\large $a$};
	\node at (6.75,-1) {\large $b$};
 \end{tikzpicture}
\vspace{2em}
\caption{(Color online) Vertex corrections to the current operator $J_{\alpha}(\mathbf{k})$ by the ladder sum. $j_{\alpha}(\mathbf{k})$ is the bare current operator and each ladder is given by the disorder strength $n_i u_0^2$. $a$, $b$ are labels of spinor indices.}
\label{fig:ladder}
\end{figure*}

In this section, we calculate the diffusive electrical transport by a diagrammatic approach, which is similar to the case of the two dimensional semi-Dirac spectrum\cite{Adroguer2015}. To lowest order in the disorder strength, the self-energy is given by
\begin{equation}
\hat{\Sigma }(\mathbf{k}, \omega )=\int \frac{d^3\mathbf{k}'}{(2\pi )^3}\overline{V\left(\mathbf{k}'\right)V\left(-\mathbf{k}'\right)}\hat{G}\left(\mathbf{k}-\mathbf{k}',\omega
\right),
\end{equation}
where $V\left(\mathbf{k}\right)$ is the Fourier transform of disorder potential. The over line denotes a statistical average over realizations of the random potential and
\begin{equation}
\hat{G}(\mathbf{k}, \epsilon_F)=\frac{-\epsilon_F -\frac{k_x^2-k_y^2}{m}\sigma _x-\frac{2k_xk_y}{m}\sigma _y-v_zk_z\sigma _z}{-\epsilon_F^2+\epsilon(\mathbf{k})^2},
\end{equation}
is the noninteracting Green's function in spinor representation. The bare current operators are, 
\begin{eqnarray}
j_x&=&2\left(\frac{k_x}{m}\sigma_x+\frac{k_y}{m}\sigma _y\right), \nonumber \\
j_y&=&-2\left(\frac{k_y}{m}\sigma_x-\frac{k_x}{m}\sigma _y\right), \nonumber \\
j_z&=&v_z \sigma_z.
\end{eqnarray}
The imaginary part of the retarded/advanced self-energy is obtained as,
\begin{eqnarray}
\text{Im}\hat{\Sigma }^{R/A}(\mathbf{k}, \epsilon_F)&=&\frac{u_0^2n_i \pi ^2}{2(2\pi )^3}\frac{m}{v_z} \times \nonumber \\
&&\int d\epsilon ^2\text{Im}\frac{\left(-\epsilon_F \mp
\text{i0}^+\right)}{\epsilon ^2-\epsilon_F^2\mp 2 \epsilon_F i 0^+} \nonumber \\
&=&\mp u_0^2n_i\frac{1}{16}\frac{m}{v_z}|\epsilon_F|.
\end{eqnarray}
As $j_x$ and $j_y$ are odd in momentum, there are no vertex corrections in the $x$ and $y$ directions:
\begin{equation}
J_x=j_x, J_y=j_y.
\end{equation}
Thus, the in-plane longitudinal electrical conductivity at zero temperature is evaluated as\cite{Datta1997}
\begin{eqnarray}
\sigma _{x x}&=&\frac{\hbar }{2\pi  V}\text{ReTr}\left[j_x(\mathbf{k})G^R\left(\mathbf{k},\epsilon _F\right)j_x(\mathbf{k})G^A\left(\mathbf{k},\epsilon _F\right)\right] \nonumber \\
&\approx&\frac{8\hbar }{3\pi ^2 u_0^2n_im}\left|\epsilon _F\right|,
\end{eqnarray}
up to lowest order in disorder strength--in agreement with Eq.(\ref{eq:sigmadisorder}). For conductivity in the $z$-direction, we need to consider the vertex correction by the ladder sum in Fig. \ref{fig:ladder} as,
\begin{widetext}
\begin{eqnarray}
\label{eq:vertexjz}
J_z(\mathbf{k})=j_z(\mathbf{k})+J_z(\mathbf{k})\Pi (\mathbf{k}) n_iu_0^2 \nonumber \\
\Rightarrow J_z(\mathbf{k})=j_z(\mathbf{k})\left(I\otimes I-n_iu_0^2\Pi (\mathbf{k})\right)^{-1}
\end{eqnarray}
where
\begin{eqnarray}
\Pi \left(\epsilon _F\right)&=&\int \frac{d^3\mathbf{k}}{(2\pi )^3}G^R\left(\mathbf{k},\epsilon _F\right)\otimes G^A\left(\mathbf{k},\epsilon _F\right)^T \nonumber \\
&\simeq& \int \frac{k_{\perp}dk_{\perp}dk_zd\alpha }{(2\pi )^3}\pi \frac{\delta \left(\epsilon ^2-\epsilon _F^2+\left(\frac{u_0^2n_im}{16v_z}\right)^2\epsilon
_F^2\right)}{2\frac{u_0^2n_im}{16v_z}\epsilon _F^2} \times \nonumber \\
&&\left[\epsilon _F^2\left(1+\left(\frac{u_0^2n_i m}{16v_z}\right)^2\right)I\otimes I+\frac{k_{\perp}^4}{m^2}\cos
(2\alpha )^2\sigma _x\otimes \sigma _x-\frac{4k_{\perp}^4\cos (\alpha )^2\sin (\alpha )^2}{m^2}\sigma _y\otimes \sigma _y+\left(v_zk_z\right)^2\sigma
_z\otimes \sigma _z\right] \nonumber \\
&=&\frac{1}{2u_0^2n_i}\left[ \left(1+\left(\frac{u_0^2n_im}{16v_z}\right)^2\right)I\otimes I+\frac{1}{4}\left(1-\left(\frac{u_0^2n_im}{16v_z}\right)^2\right)\left(\sigma
_x\otimes \sigma _x-\sigma _y\otimes \sigma _y\right)+\frac{1}{2} \left(1-\left(\frac{u_0^2n_im}{16v_z}\right)^2\right)\sigma _z\otimes \sigma
_z\right]
\end{eqnarray}
Then
\begin{equation}
\label{eq:vertex}
\left[I\otimes I-n_iu_0^2\Pi (\mathbf{k})\right]^{-1}\cdot \sigma _z=\frac{1}{1- \frac{1}{2}\left(1+\left(\frac{u_0^2n_im}{16v_z}\right)^2\right)}\sigma_z.
\end{equation}
Pluggin Eq.(\ref{eq:vertex}) back into Eq.(\ref{eq:vertexjz}), we find, 
\begin{equation}
J_z(\mathbf{k})=\frac{1}{1- \frac{1}{2}\left(1+\left(\frac{u_0^2n_im}{16v_z}\right)^2\right)}v_z\sigma _z.
\end{equation}
The longitudinal electrical conductivity in $z$ direction can be evaluated as
\begin{eqnarray}
\sigma _{z z}&=&\frac{\hbar }{2\pi }\text{Tr}\left[J_z\cdot \Pi \cdot j_z\right]=\frac{\hbar }{2\pi u_0^2n_i}\text{Tr}\left[\left(J_z-j_z\right)\cdot
j_z\right] \nonumber \\
&=&\frac{\hbar  v_z^2}{\pi u_0^2n_i}\frac{ \frac{1}{2}\left(1+\left(\frac{u_0^2n_im}{16v_z}\right)^2\right)}{1- \frac{1}{2}\left(1+\left(\frac{u_0^2n_im}{16v_z}\right)^2\right)} \nonumber \\
&\approx& \frac{\hbar  v_z^2}{\pi u_0^2n_i},
\end{eqnarray}
which is also in agreement with Eq.(\ref{eq:sigmadisorder}).
\end{widetext}


\end{document}